\documentclass[12pt,english]{iopart}

\usepackage{iopams} 

\usepackage{bbm}
\usepackage{color}
\usepackage{graphicx}
\usepackage{morefloats}
\usepackage{slashed}

\def\blr#1{\left(#1\right)}

\def\eq#1{(\ref{#1})}
\def\Eq#1{Eq.~(\ref{#1})}
\newcommand{\be}{\begin{equation}}
\newcommand{\ee}{\end{equation}}
\newcommand{\bea}{\begin{eqnarray}}
\newcommand{\eea}{\end{eqnarray}}
\newcommand{\bse}{\begin{subequations}}
\newcommand{\ese}{\end{subequations}}
\def\Fig#1{Fig.~\ref{#1}}
\def\im{\mrm{i}}

\def\mc#1{\mathcal{#1}}
\def\mrm#1{\mathrm{#1}}
\def\N0{\mathbbm N_0}
\def\Nc{N_\mrm{c}}

\def\non{\nonumber}

\def\pd{\partial}
\def\Phib{\bar{\Phi}}
\def\Phir{\Phi_\mrm{r}}
\def\Phii{\Phi_\mrm{i}}

\def\qqb{\mrm{q\bar{q}}}
\def\Ref#1{Ref.~\cite{#1}}
\def\Refs#1{Refs.~\cite{#1}}
\def\Sec#1{Sec.~\ref{#1}}

\newcommand{\sns}{\sigma_\mrm{l}}
\newcommand{\sy}{\sigma_\mrm{\!s}}
\def\Tab#1{Table~\ref{#1}}

\graphicspath{{./figures/}}

\definecolor{orange}{rgb}{1,0.5,0}

\begin{document}

\title[PQM model at non-vanishing density with the unquenched Polyakov-loop potential]{Phase diagram and nucleation in the Polyakov-loop--extended Quark-Meson truncation of QCD with the unquenched Polyakov-loop potential}

\author{Rainer Stiele}
\address{Univ.~Lyon, Universit\'e Claude Bernard Lyon 1, CNRS/IN2P3, IPN-Lyon, F-69622, Villeurbanne, France}
\address{Institut f\"{u}r Theoretische Physik, Goethe-Universit\"at Frankfurt, Max-von-Laue-Stra\ss e 1, 60438 Frankfurt am Main, Germany}
\eads{\mailto{r.stiele@ipnl.in2p3.fr}}

\author{J\"urgen Schaffner-Bielich}
\address{Institut f\"{u}r Theoretische Physik, Goethe-Universit\"at Frankfurt, Max-von-Laue-Stra\ss e 1, 60438 Frankfurt am Main, Germany}
\eads{\mailto{schaffner@astro.uni-frankfurt.de}}

\maketitle

\begin{abstract}
	The unquenching of the Polyakov-loop potential showed to be an important improvement for the description of the phase structure and thermodynamics of strongly-interacting matter at zero quark chemical potentials with Polyakov-loop--extended chiral models.
	This work constitutes the first application of the quark backreaction on the Polyakov-loop potential at non-zero density. The observation is that it links the chiral and deconfinement phase transitions also at small temperatures and large quark chemical potentials.
	The build-up of the surface tension in the Polyakov-loop--extended Quark-Meson model is explored by investigating the two and 2+1--flavour Quark-Meson model and analysing the impact of the Polyakov-loop extension. In general, the order of magnitude of the surface tension is given by the chiral phase transition. The coupling of the chiral and deconfinement transitions with the unquenched Polyakov-loop potential leads to the fact that the Polyakov loop contributes at all temperatures.
\end{abstract}

%s%s%s%s%s%s%s%
\section{Introduction}\label{sec:Intro}
%s%s%s%s%s%s%s%

The extended mean-field version of the Polyakov-loop--extended Quark-Meson (PQM) model and the no-sea approximation of the Polyakov-Nambu-Jona-Lasinio model showed to be capable to reproduce the crossover at vanishing densities of the strong interaction as seen in lattice calculations.
The important ingredients  to achieve this agreement proved to be the enhancement of the Polyakov-loop potential from a pure gauge potential to the unquenched glue potential that includes backreaction effects from quarks and including thermal fluctuations of mesons \cite{Haas:2013qwp,Herbst:2013ufa,Torres-Rincon:2016ahl}. The lattice data were used to adjusted the free parameters, which are the critical scale of the Polyakov-loop potential and the mass of the $\sigma$-meson. In \Ref{Stiele:2013pma} we have applied this framework to investigate the phase structure at non-vanishing isospin before the onset of pion condensation. The present work presents the first application of the unquenched Polyakov-loop potential at non-zero net quark density and explores its applicability \footnote{During finalising this manuscript a similar application appeared in arXiv \cite{Kovacs:2016juc}.}
.
The observation is that it links the chiral and deconfinement phase transitions also at small temperatures and large quark chemical potentials. This is a feature that is also seen in the functional renormalization group (FRG) improvement of the PQM model when using the Yang-Mills Polyakov-loop potential \cite{Herbst:2010rf,Herbst:2013ail}. Therefore, the further investigation is interesting although mesonic fluctuations still have to be implemented to the equations of motion in a next step. For now results within the extended mean-field version including thermal fluctuations of mesons to thermodynamics will be presented.
The effect of unquenching the Polyakov-loop potential is a shift of a hypothetical critical endpoint towards smaller temperatures at similar chemical potential. Furthermore, there is a minor decrease in size of the spinodal region.
Moreover, we discuss the build-up of the surface tension in the PQM model by investigating the two and 2+1--flavour QM model and analysing the impact of the Polyakov-loop extension. In general, the order of magnitude of the surface tension is given by the chiral phase transition. The coupling of the chiral and deconfinement transitions with the unquenched Polyakov-loop potential leads to the fact that the Polyakov loop contributes at all temperatures.

In the next section the theoretical framework that is used is explained. Section \ref{sec:Resultsmu0} summarises the results at vanishing quark chemical potentials before exploring the impact of unquenching the Polyakov-loop potential on the phase structure at non-zero quark chemical potentials in \Sec{sec:ResultsMuQ}.
The build-up of the surface tension in the QM framework and the impact of the Polyakov-loop extension are discussed in \Sec{sec:Nucleation}.
Finally, the findings are summarised in \Sec{sec:Conclusions}.

%s%s%s%s%s%s%s%
\section{Theoretical framework}\label{sec:Framework}
%s%s%s%s%s%s%s%

We perform our investigation within the 2+1-flavour extended mean-field Polyakov-loop--extended Quark-Meson model enhanced by the meson contributions to thermodynamics and with an unquenched Polyakov-loop potential.\footnote{The extended mean-field Polyakov-loop Quark-Meson model does not consider fluctuations of mesons but only thermal and quantum fluctuations of the quarks \cite{Schaefer:2007pw} as we discuss in the course of this section. To achieve a better description of thermodynamics in the phase where chiral symmetry is broken, we add the thermal fluctuations of mesons to thermodynamics as we did in \Refs{Haas:2013qwp,Stiele:2013pma,Herbst:2013ufa} and as we discuss in \Sec{sec:Resultsmu0}.}
This framework describes two important properties of QCD which are chiral symmetry and centre symmetry.
While spontaneous breaking of the former gives rise to constituent quark masses, spontaneous breaking of center symmetry indicates deconfinement.
These properties are described by the effective Lagrangian
\bea
	\mc{L} &=& \bar{q} \blr{i\slashed{D} - g \,\phi_5 + \gamma_0 \,\mu_f} q + \Tr\blr{\pd_\mu \phi^\dagger \,\pd^\mu\phi} -\nonumber\\
	&&- \,m^2\, \Tr \blr{\phi^\dagger\phi} - \lambda_1 \left[ \Tr\blr{\phi^\dagger\phi} \right]^2 - \lambda_2\, \Tr \blr{\phi^\dagger\phi}^2 + \nonumber\\
	&&+\, c \blr{\det\phi + \det\phi^\dagger} + \Tr \left[ H \blr{\phi + \phi^\dagger} \right] - \nonumber\\
	&& -\, \mc{U}\! \blr{\Phir,\Phii}\;,
	\label{eq:Lagrangian}
\eea
where $\phi$ and $\phi_5$ are 3$\times$3 matrices that combine scalar and pseudoscalar meson fields \cite{Lenaghan:2000ey} and $\mc{U}$ is the potential of the Polyakov loop $\Phi$ \cite{Meisinger:1995ih,Pisarski:2000eq,Scavenius:2002ru,Mocsy:2003qw,Fukushima:2003fw,Megias:2004hj,Ratti:2005jh}, which is the trace of a Wilson loop in the temporal direction over the temporal component of the gauge field $A_\mu$ and as such in general is a complex scalar field $\Phi = \Phir + \im \Phii$ \cite{Polyakov:1978vu}.

The corresponding partition function is a path integral over all occurring fields of the in-medium effective action. In a mean-field analysis, the field is replaced by its spatially and temporally constant background such that the path integral turns into an ordinary integral over that field.
Furthermore, the integral is trivially restricted to that mean-field that contributes the most to the partition function and thus minimises the effective action or potential.
This is how the meson fields and gauge field are treated. Replacing these fields with their expectation values their derivative terms in the Lagrangian \eq{eq:Lagrangian} vanish.
The remaining integration over the fermions can be performed following the standard derivation \cite{Kapusta:2006pm} as Gaussian integral over Grassmann fields followed by applying the Matsubara formalism.
The expression of the grand-canonical potential then takes the form
\bea
	\hskip-10ex \Omega\blr{\sns,\sy,\Phir,\Phii;\,T,\mu_f} &=& U\blr{\sns,\sy} + \mc{U}\blr{\Phir,\Phii;\,T} + \Omega_\qqb\blr{\sns,\sy,\Phir,\Phii;\,T, \mu_f} \;. \qquad
	\label{eq:grand_canon_pot}
\eea
It is a function of the order parameters for center symmetry and for chiral symmetry in the light and strange quark sectors and depends on the temperature and the quark chemical potentials.

The self-interaction potential of the meson fields $U\!\blr{\sns,\sy}$ is that of the isospin-symmetric linear sigma model \cite{Lenaghan:2000ey,Schaefer:2008hk}.
It describes spontaneous and explicit breaking of chiral symmetry in the light and strange quark sector. For an extension that distinguishes between up and down quark condensates at non-zero isospin, see e.g.~\Ref{Stiele:2013pma}. We use the same form and phenomenological input as in our previous work in \Refs{Mintz:2012mz,Haas:2013qwp,Stiele:2013pma,Beisitzer:2014kea,Zacchi:2015lwa} which is specified in \Tab{tab:chiral_pot_constants}. For the mass of the scalar $\sigma$-meson we use $m_\sigma=400\,\mrm{MeV}$ to obtain best agreement with lattice results at vanishing density as will be shown in \Sec{sec:Resultsmu0}. Note that even in the vacuum, not only the mesonic contribution $U\!\blr{\sns,\sy}$ but also the contribution of the quark quantum-fluctuations $\Omega_\qqb^\mrm{vac}\blr{\sns,\sy}$ which will be introduced in the next paragraph contribute to the meson masses.% \new{which is in agreement with the Particle Data Group \cite{Agashe:2014kda} values. For the mass of the scalar $\sigma$-meson we use $m_\sigma=400\,\mrm{MeV}$ to obtain best agreement with lattice results at vanishing density as will be shown in \Sec{sec:Resultsmu0}.}

%%%%%%%%%%%%%%%%%%%%%%%%%%%%%%%%%%%%%%%%%%
\begin{table}
	\caption{Values of decay constants of pseudoscalar mesons and meson masses in the vacuum in accordance to Ref.~\cite{Agashe:2014kda}, to which the parameters of the mesonic potential are adjusted, as well as the chosen value of the constituent quark mass of the light (up and down) quarks that is used to fix the quark-meson Yukawa coupling.}
	\begin{center}
		\begin{tabular}{lcccccccr}
			\hline\hline\vspace{-0.4cm}\\
                  			Constant & $f_\pi$ & $f_\mrm{K}$ & $m_\pi$ & $m_\mrm{K}$ & $m_\eta$ & $m_{\eta'}$ & $m_\sigma$ & $m_\mrm{u,d}$\\
			\vspace{-0.4cm}\\\hline\vspace{-0.4cm}\\
                  			Value [MeV] & 92 & 110 & 138 & 495 & 548 & 958 & 400 & 300\\\vspace{-0.4cm}\\
			\hline\hline		
		\end{tabular}
	\end{center} 
	\label{tab:chiral_pot_constants}
\end{table}
%%%%%%%%%%%%%%%%%%%%%%%%%%%%%%%%%%%%%%%%%%
%t%t%t%
\begin{table}
	\caption{Values for the remaining scalar meson masses in the vacuum and the constituent mass of strange quarks in units of MeV resulting from the input in \Tab{tab:chiral_pot_constants}. For a comparison to experimental results see e.g.~\Ref{Lenaghan:2000ey}}
	\begin{center}
		\begin{tabular}{cccr}
			\hline\hline\vspace{-0.4cm}\\
			$m_{a_0}$ & $m_\kappa$ & $m_{f_0\blr{1370}}$ & $m_\mrm{s}$\\
			\vspace{-0.4cm}\\\hline\vspace{-0.4cm}\\
		 	1122	 & 1183 & 1204 & 417 \\\vspace{-0.4cm}\\
			\hline\hline		
		\end{tabular}
	\end{center} 
	\label{tab:RemainingMesonMasses}
\end{table}
%t%t%t%

The fluctuation contribution of quarks and anti-quarks $\Omega_\qqb\blr{\sns,\sy,\Phir,\Phii;\,T, \mu_f}$, which comes from the fermionic determinant, includes the coupling to and between the Polyakov-loop variable and meson fields. It consists of two terms. 
One contribution, $\Omega_\qqb^\mrm{vac}\blr{\sns,\sy}$, results from the negative-energy states of the Dirac sea and is ultraviolet divergent. In the standard, no-sea mean-field approximation it is neglected by saying that it can be absorbed in the renormalisation of the vacuum since it has no explicit dependence on temperature and quark chemical potentials \cite{Scavenius:2000qd}.
However, it is implicitly medium dependent via the quark condensates and can be normalised, e.g.~in the dimensional regularisation scheme yielding a logarithmic correction \cite{Skokov:2010sf} that is taken into account in the extended mean-field analysis \cite{Gupta:2011ez,Chatterjee:2011jd,Schaefer:2011ex}.
This term contributes besides the mesonic potential $U\!\blr{\sns,\sy}$ to the vacuum properties, so at $T=\mu_f=0$. Therefore, it has to be considered in the adjustment of the parameters of the mesonic potential to the physical meson masses and decay constants in the vacuum which are given in \Tab{tab:chiral_pot_constants}. The dependence of $\Omega_\qqb^\mrm{vac}\blr{\sns,\sy}$ on its regularisation scale cancels neatly with that of these parameters as is shown in detail in \Ref{Chatterjee:2011jd}. The remaining scalar meson masses and the constituent strange quark mass resulting from the input in \Tab{tab:chiral_pot_constants} are given in \Tab{tab:RemainingMesonMasses}. Note that only the mass of the $f_0\blr{1370}$-meson depends on the value used for the mass of the $\sigma$-meson. How these results compare to experimental values is e.g.~discussed in \Ref{Lenaghan:2000ey} and the medium dependence of all meson masses is e.g.~discussed in \Ref{Schaefer:2008hk}.\\
The remaining part of $\Omega_\qqb$ is the kinetic quark-antiquark contribution $\Omega_\qqb^\mrm{th}\blr{\sns,\sy,\Phir,\Phii;\,T, \mu_f}$ that depends on the thermodynamic variables.
At non-vanishing quark chemical potential it contributes an imaginary part to the effective potential, $\Omega_\qqb^\mrm{th} = \Omega_\qqb^\mrm{R} + \im\,\Omega_\qqb^\mrm{I}$. This is the manifestation of the fermion sign problem in the Polyakov-loop extensions of the Quark-Meson model and NJL model \cite{Fukushima:2006uv,Rossner:2007ik,Mintz:2012mz}.
The common approach to circumvent this sign problem is to redefine the Polyakov loop $\Phi$ and its complex conjugate $\Phib$ as two independent, real variables, see e.g.~\Ref{Ratti:2005jh}. But by this approach, the state of thermodynamical equilibrium is identified only with a saddle point but not with a minimum of the effective potential. But one can only calculate quasi-equilibrium properties of the system, such as the surface tension and nucleation rate in a first-order phase transition as we want to do here with equilibrium states described by minima of the effective potential. 
As we have shown in \Ref{Mintz:2012mz} in accordance with \Ref{Rossner:2007ik} another way to avoid the sign problem is to neglect the imaginary part of the effective potential as a lowest-order perturbative approximation \cite{Rossner:2007ik,Mintz:2012mz}. Dropping the complex part of the effective potential implies that the imaginary part of the Polyakov loop $\Phii$ is zero but this approach has the advantage that the state of equilibrium is a minimum of the effective potential.
Using the phase reweighting method to obtain $\Phii\neq0$ as an effect of the imaginary part of the potential beyond lowest order as in the dense-heavy model in \Ref{Fukushima:2006uv} remains for future work.

The potential of the Polyakov loop, $\mc{U}\!\blr{\Phir,\Phii;T}$ should mimic a background of gluons and controls the dynamics of the Polyakov loop.
A common, simple way to obtain an effective Polyakov-loop potential is to construct a potential that respects all given symmetries and contains the spontaneous breaking of Z(3) symmetry if the system is in the deconfined phase \cite{Svetitsky:1982gs,Svetitsky:1985ye,Banks:1983me}.
A polynomial forms in this sense the minimal content of a Polyakov-loop potential \cite{Pisarski:2000eq,Ratti:2005jh}.
The ansatz for the Polyakov-loop potential can be enhanced by including the term that arises if one integrates out the SU(3) group volume in the generating functional for the Euclidean action. This integration can be performed via the so-called Haar measure and takes the form of a Jacobian determinant. Its logarithm adds as an effective potential to the action in the generating functional \cite{Fukushima:2003fw,Roessner:2006xn}. Reference \cite{Lo:2013hla} went beyond a minimal content for the Polyakov-loop potential and kept the higher-order terms of the polynomial parametrisation of the Polyakov-loop potential and added the logarithmic term to consider the group volume additionally.
On the parameters of the parametrisations some general constraints can be imposed, especially by applying and restricting the Polyakov-loop potential to pure gauge (Yang-Mills) theory.
The remaining open parameters are determined in \Refs{Scavenius:2002ru,Ratti:2005jh,Roessner:2006xn,Lo:2013hla} by fitting both the lattice data for pressure, entropy density and energy density and the evolution of the Polyakov loop $\langle \Phi \rangle$ on the lattice in pure gauge theory. Reference~\cite{Lo:2013hla} adjusted their parameters in addition to lattice data of the longitudinal and transverse Polyakov-loop susceptibilities.
The different parametrisations that we apply are summarised together with their parameter sets in \ref{app:PloopPots} within a more detailed discussion of the Polyakov-loop potential.\\
This construction of the Polyakov-loop potential and the fitting of its parameters entails that it models the pure gauge potential $\mc{U}_\mrm{YM}\blr{t_\mrm{YM}}/T^4$. The transition scale of the Polyakov-loop potential is accordingly the critical temperature of pure gauge theory, $T_0^\mrm{YM}=270\,\mrm{MeV}$.
However, in full dynamical QCD, one important effect of fermionic matter fields is to change the scale $\Lambda_\mrm{QCD}$ to which the transition temperature of the Polyakov-loop potential is linked.
To consider this aspect of the backreaction of quarks to the gauge sector, \Ref{Schaefer:2007pw} estimated the running coupling of QCD by consistency with hard thermal loop perturbation theory calculations \cite{Braun:2005uj,Braun:2006jd} and they mapped the effect to an $N_f$-dependent modification of the expansion coefficients of the Polyakov-loop potential. Their result is a $N_f$-dependent decrease of $T_0$.
This accounts partially for the unquenching of the pure gauge Polyakov-loop potential to an effective glue potential in QCD.
Besides the flavour dependency of the transition scale of the glue sector, one can consider its dependence on the quark density. Such a dependency has to be expected in view of a $\mu_\mrm{q}$-dependent colour screening effect due to quarks. In \Ref{Schaefer:2007pw} a $\mu_\mrm{q}$-dependent small correction to the running coupling was motivated by using HTL/HDL (hard-thermal/dense-loop) theory and by comparison to the one found in FRG calculations \cite{Braun:2009gm}. The description presented in \Refs{Schaefer:2007pw,Herbst:2013ail} can be generalised to allow for different chemical potentials for each quark flavour \cite{Stiele:2013pma}.

Still, the Polyakov-loop potential is an approximation to the Yang-Mills glue potential. However, in Polyakov-loop--extended models for (full) QCD, this Polyakov-loop potential has to be replaced by the QCD glue potential that takes into account the backreaction of quarks into the Polyakov-loop effective potential.
Using the functional renormalisation group (FRG) approach, \Refs{Braun:2007bx,Braun:2010cy,Fister:2013bh} calculated the non-perturbative Polyakov-loop potential of pure gauge theory and \Refs{Braun:2009gm,Pawlowski:2010ht} calculated the QCD analogue taking into account the backreaction of the quark degrees of freedom on the gluon propagators. The latter includes the quark part of the gluonic vacuum polarisation but does not include the fermionic part of the full QCD potential.
As we discussed in detail in \Ref{Haas:2013qwp} one observes for a given non-zero reduced temperature that the shapes of the potentials are very similar but that there is a shift between both potentials.
This observation can be exploited to estimate how to convert the Yang-Mills potential for the Polyakov loop to a glue potential in full QCD that contains backreaction effects from quarks,
%e%e%e%
\bea
	\frac{\mc{U_\mrm{glue}}}{T^4} \blr{\Phir,\Phii,t_\mrm{glue}} &=& \frac{\mc{U_\mrm{YM}}}{T_\mrm{YM}^4}\blr{\Phir,\Phii,t_\mrm{YM}(t_{\rm glue})} \;,
	\label{eq:gluePotYMPot}
\eea
%e%e%e%
by relating the reduced temperature scales of both potentials
%e%e%e%
\be
	t_{\mrm{glue}}=\frac{T-T^\mrm{glue}_\mrm{cr}}{T^\mrm{glue}_\mrm{cr}} \qquad \mrm{and} \qquad t_\mrm{YM}=\frac{T_\mrm{YM}-T^\mrm{YM}_\mrm{cr}}{T^\mrm{YM}_\mrm{cr}}\;.
	\label{eq:reducedTemps}
\ee
%e%e%e%
The comparison of the potentials yields that the relation between the two temperature scales
%e%e%e%
\be
	t_\mrm{YM} (t_\mrm{glue}) \approx 0.57\, t_\mrm{glue}\;
	\label{eq:tYMtglue}
\ee
%e%e%e%
minimises the difference between both potentials, the pure gauge Polyakov-loop potential and unquenched glue potential. Limitations of this approximation and details of its derivation are given in \Ref{Haas:2013qwp}. In this work, we will apply for the first time this unquenching of the Polyakov-loop potential at non-vanishing quark density and discuss its implications.

%s%s%s%s%s%s%s%
\section{Order Parameters and Thermodynamics at zero Chemical Potentials}\label{sec:Resultsmu0}
%s%s%s%s%s%s%s%

\begin{figure}[b]
	\centering
	\includegraphics[width=.49\textwidth]{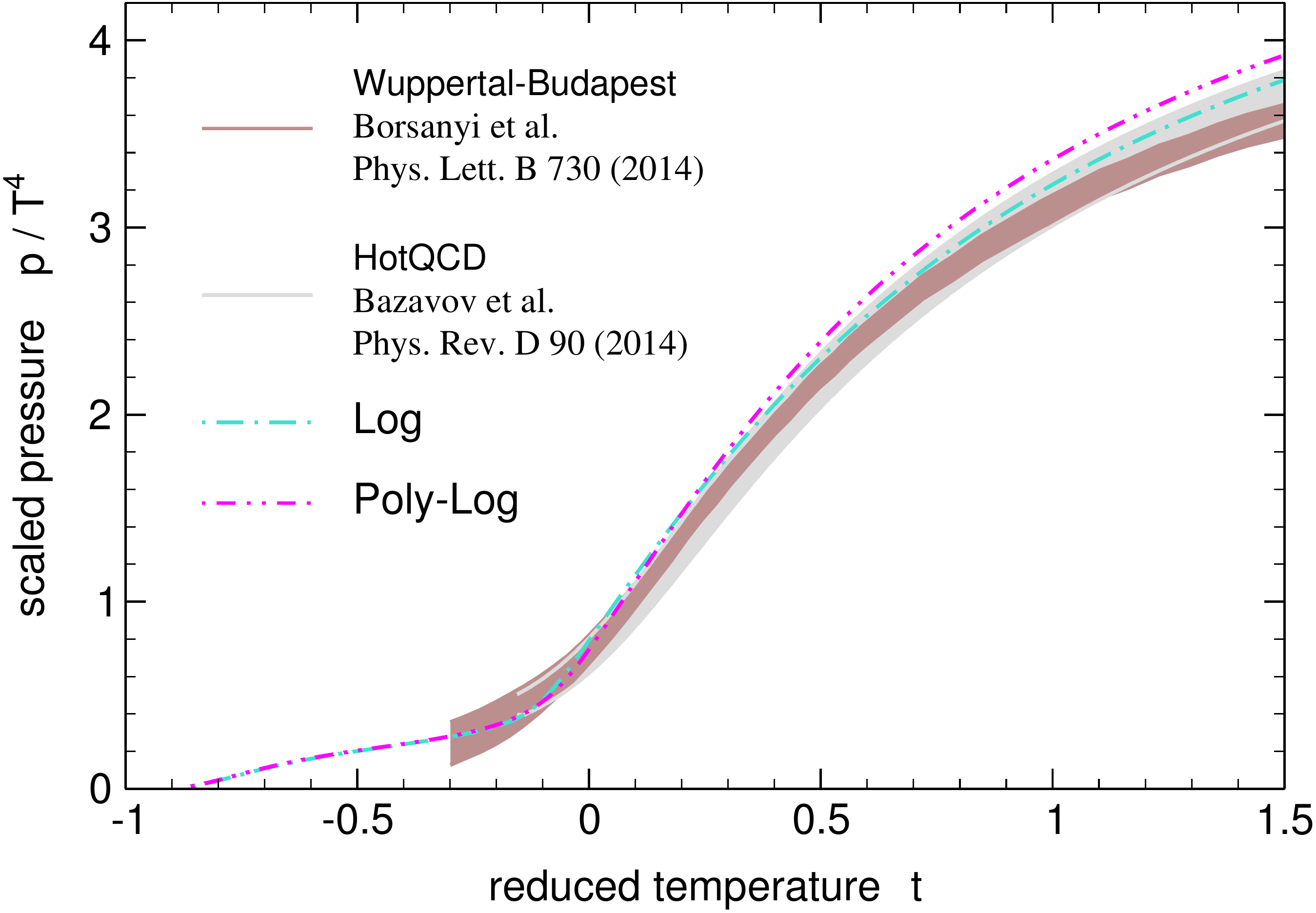}
	\hfill
	\includegraphics[width=0.49\textwidth]{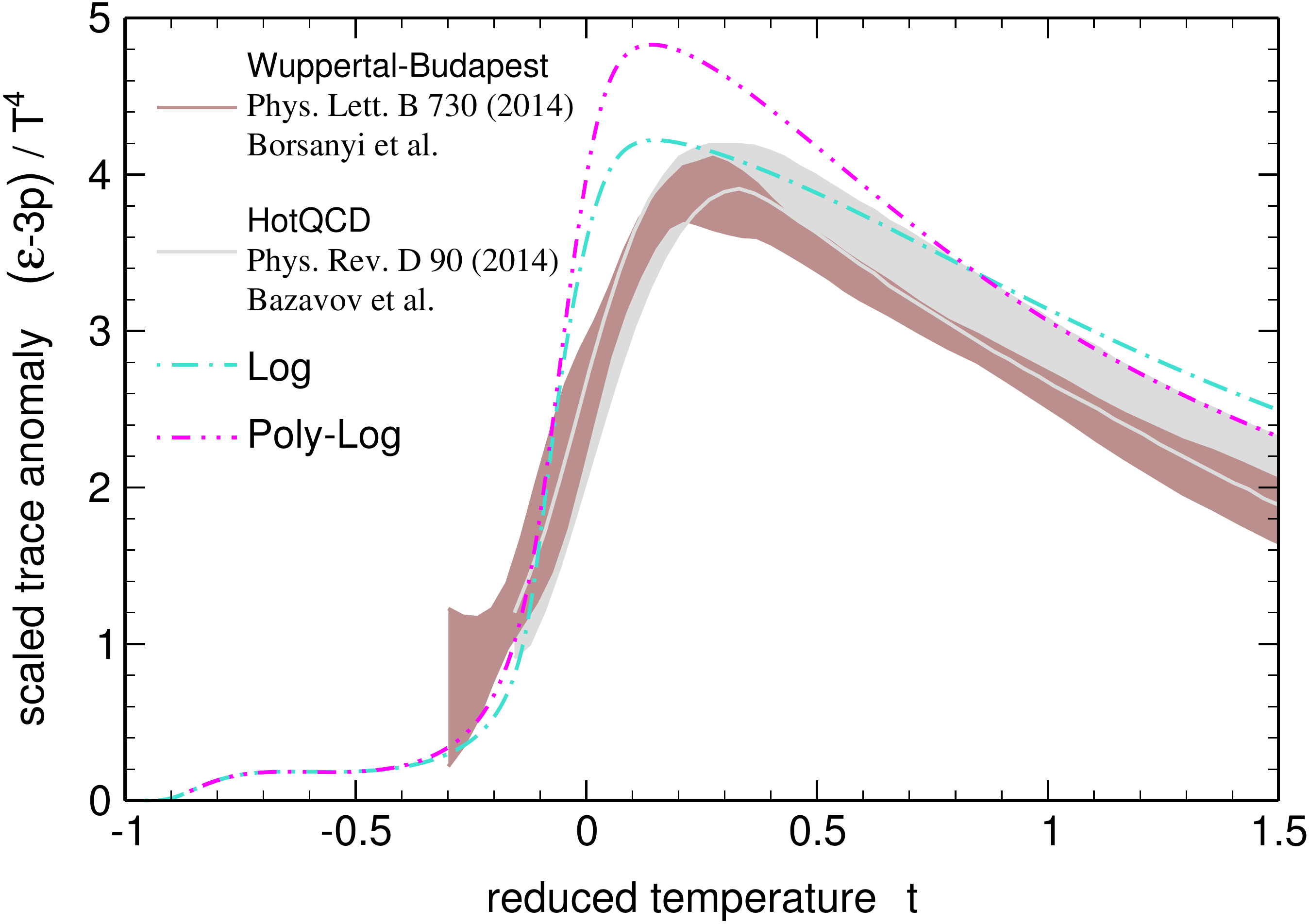}
	\caption{Results for the scaled pressure $p/T^4$ (left) and trace anomaly $\blr{\epsilon-3p}/T^4$ (right) as a function of temperature at $\mu_f=0$ for the different parametrisations of the Polyakov-loop potential.
	The results are compared to the lattice calculation of \Refs{Borsanyi:2013bia,Bazavov:2014pvz}.}
	\label{fig:thd_params_best}
\end{figure}

\begin{figure}%[!hbt]
	\centering
	\includegraphics[width=0.49\textwidth]{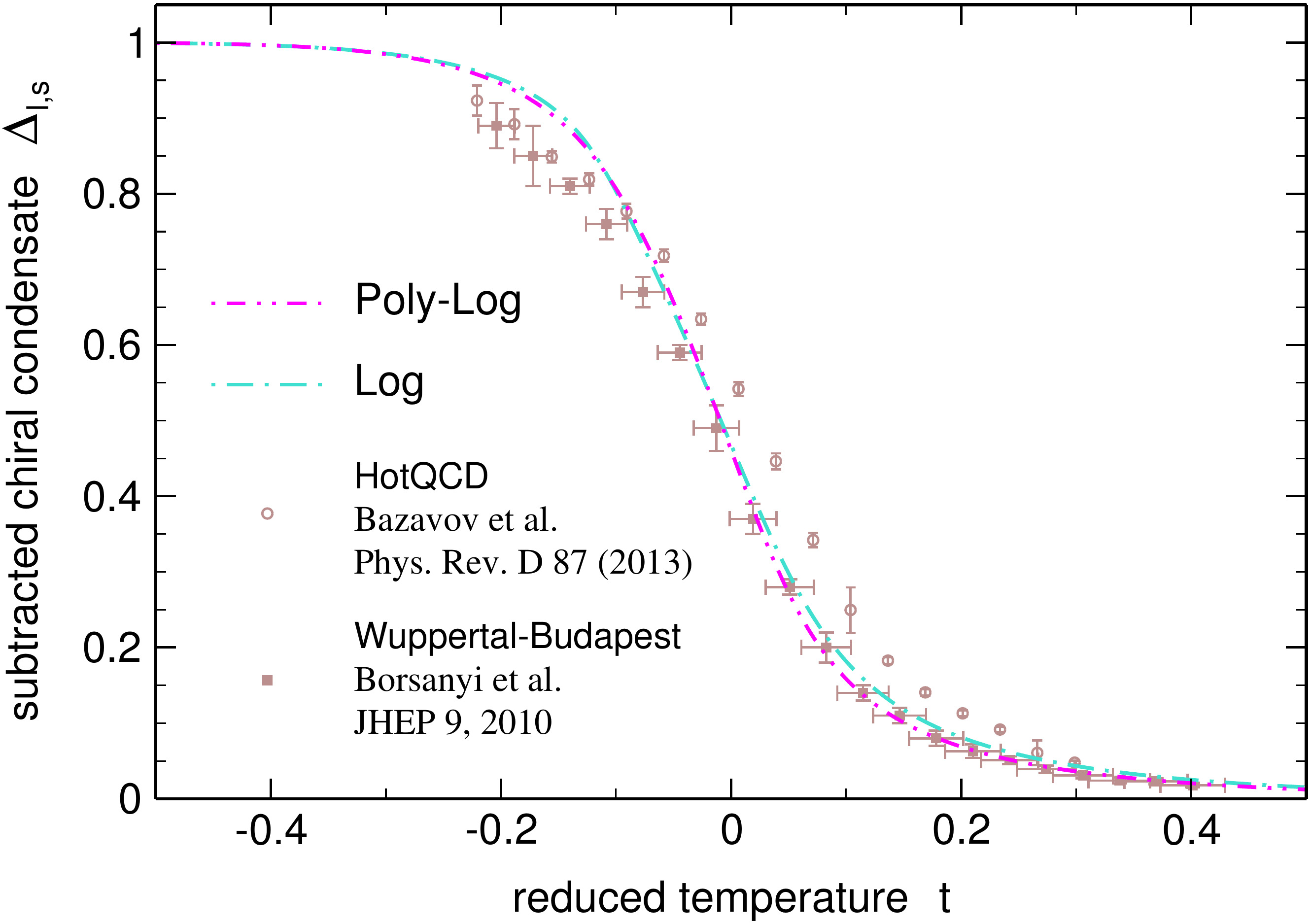}
	\hfill
	\includegraphics[width=0.49\textwidth]{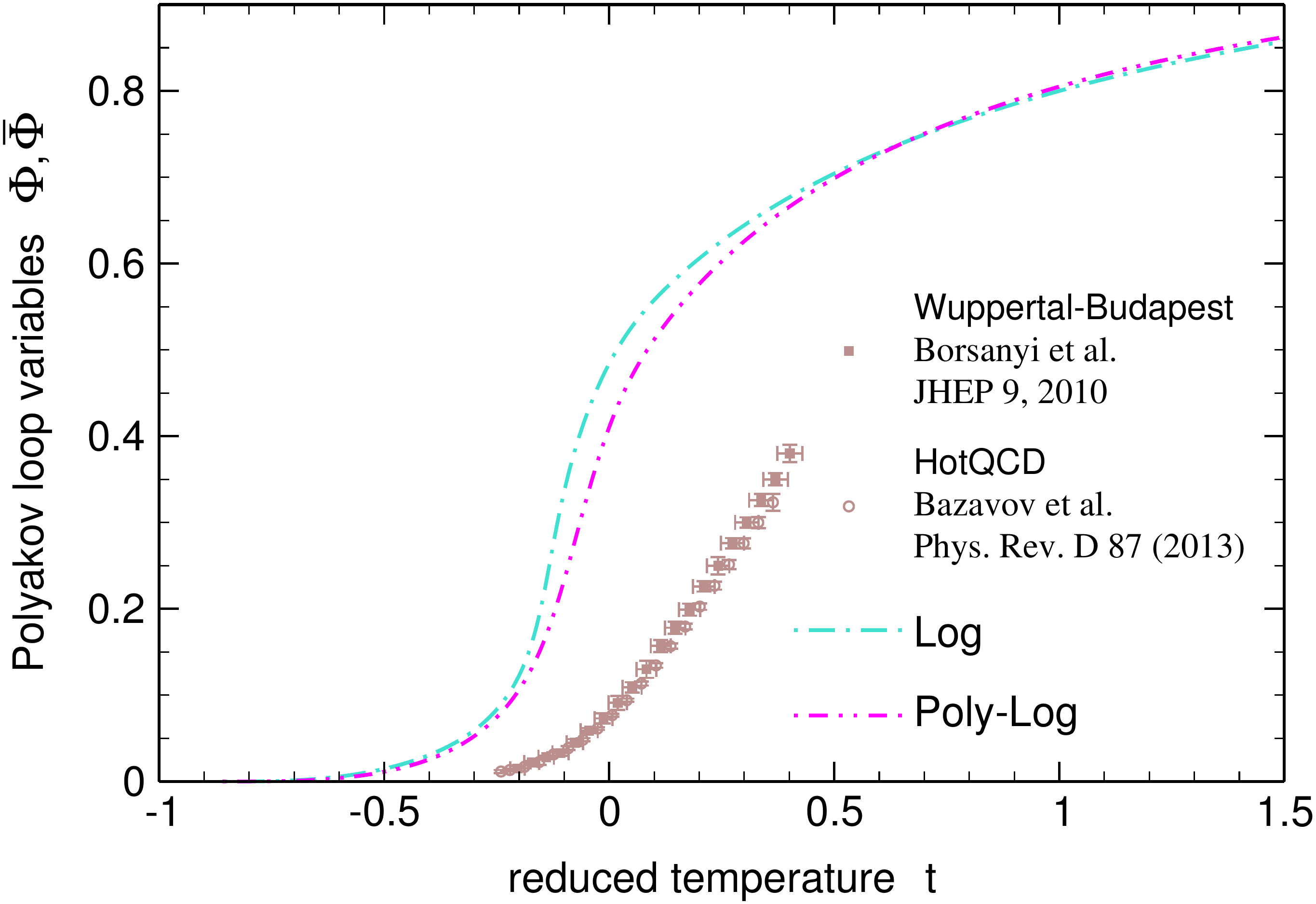}
	\caption{Results for the subtracted chiral condensate $\Delta_\mrm{l,s}$ (left) and the Polyakov loop (right) as a function of temperature at $\mu_f=0$ for the different parametrisations of the Polyakov-loop potential. The results are compared to the lattice calculations of \Refs{Borsanyi:2010bp,Bazavov:2013yv}.}
	\label{fig:ops_params_best}
\end{figure}

Figures \ref{fig:thd_params_best} and \ref{fig:ops_params_best} show those results for thermodynamics and order parameters that are the closest to the results of lattice calculations.
%f%f%f%
%f%f%f%
%f%f%f%
%f%f%f%
The results for the chiral order parameter and thermodynamics either agree quantitatively with the lattice results or are at least within the trend of the data.
A big difference is seen in the Polyakov-loop expectation value. The lattice data shows a smoother transition with significant smaller values.
However, at least a part of this discrepancy originates in the inherent approximations which are still present: the derivation of the Polyakov--Quark-Meson model entails that the Polyakov-loop variable in the thermal fermionic determinant is $\Phi\left[\langle{A_0}\rangle\right]$ and not $\langle\Phi\left[A_0\right]\rangle$ as used in the Polyakov-loop potentials $\mc{U}$ and computed on the lattice. However, the continuum definition serves as an upper bound for the lattice one, $\Phi\left[\langle{A_0}\rangle\right]\ge\langle\Phi\rangle$. This issue is discussed in detail and this difference is analysed quantitatively in \Ref{Herbst:2015ona}.\\
The absciss{\ae} of the figures are in units of the reduced temperature of full QCD $t = (T-T_\mrm{c})/T_\mrm{c}$. 
This choice allows one to compare the overall shape of the observables and thereby the proper inclusion of the relevant dynamics independent of a possible mismatch of the pseudocritical temperature $T_\mrm{c}$ that is scaled out.
Table \ref{tab:Tc_params_best} summarises the pseudocritical temperatures.
%t%t%t%
\begin{table}%[!hbt]
	\caption{Pseudocritical temperatures for the results for the crossover transition at $\mu_f=0$ with the different parametrisations of the Polyakov-loop potential presented in Figs.~\ref{fig:thd_params_best} and \ref{fig:ops_params_best}.
	They are determined by the peak position of the chiral susceptibility $\pd{\Delta_\mrm{l,s}}/\pd{T}$ and of the temperature derivative of the Polyakov loop $\pd{\Phi}/\pd{T}$. $T_\mrm{cr}^\mrm{glue}=240\,\mrm{MeV}$ is used as transition temperature of the Polyakov-loop potential and $m_\sigma=400\,\mrm{MeV}$ as the vacuum mass of the $\sigma$-meson.}
	\begin{center}
		\begin{tabular}{lcc}
			\hline\hline\vspace{-0.4cm}\\
										& Log	& Poly-Log \\	
			\vspace{-0.4cm}\\\hline\vspace{-0.4cm}\\
			$T_\mrm{c}$ [MeV]				& 175	& 179 \\
			$T_\mrm{d}$ [MeV]				& 152	& 165 \\\vspace{-0.4cm}\\
			\hline\hline
		\end{tabular}
	\end{center}
	\label{tab:Tc_params_best}
\end{table}\\
%t%t%t%

To achieve a better description of the thermodynamics in the phase where chiral symmetry is broken, the thermodynamics is augmented by the contribution of a gas of thermal pions as we did in \Refs{Haas:2013qwp,Stiele:2013pma,Herbst:2013ufa}.
The contribution to the pressure of each pion species is
%e%e%e%
\bea
	\hskip-5ex p_{\pi^i} &=& \frac{1}{\blr{2\pi}^3} \int_0^\infty \mrm{d^3}k\,\frac{k^2}{3 E_{\pi^i}}\,\frac{1}{e^{(E_{\pi^i}-\mu_{\pi^i})/T}-1} \qquad \mrm{with} \qquad E_{\pi^i}=\sqrt{k^2+m_{\pi^i}^2} \qquad
	\label{eq:pPi}
\eea
%e%e%e%
and $\pi^i = \pi^0$, $\pi^+$, $\pi^-$. The total contribution of the pions to the pressure is accordingly $p_\pi=p_{\pi^0}+p_{\pi^+}+p_{\pi^-}$ and overall
%e%e%e%
\bea
	p &=& -\Omega + p_\pi\;.
	\label{eq:ptot}
\eea
%e%e%e%
For the pion masses the in-medium masses are taken. These are determined by the second derivatives of the thermodynamical potential with respect to the pseudoscalar fields,
%e%e%e%
\bea
	m_{\mrm{p},ij}^2 &=& \left.\frac{\partial^2\Omega\blr{\varphi_{\mrm{s},a},\varphi_{\mrm{p},b},\Phi,\Phib;T,\mu_f}}{\partial\varphi_{\mrm{p},i}\,\partial\varphi_{\mrm{p},j}}\right|_\mrm{min}\;.
	\label{eq:MedMesMas}
\eea
%e%e%e%
The association of the pseudoscalar fields and physical pseudoscalar mesons is such that the pion masses are given by $m_{\mrm{p},11}=m_{\mrm{p},22}=m_{\mrm{p},33}$.
%The association of the pseudoscalar fields and physical pseudoscalar mesons is such that the mass of the $\pi^+$-meson is given by $m_{\mrm{p},11}$, the mass of the $\pi^-$-meson is $m_{\mrm{p},22}$ and  the mass of the $\pi^0$-meson results from $m_{\mrm{p},33}$. %Note that the charged pions couple to isospin ($\mu_{\pi^\pm}=\pm\mu_\mrm{I}=\pm\left|\mu_\mrm{u}-\mu_\mrm{d}\right|$) and therefore, the degeneracy of their masses with that of the $\pi^0$-meson becomes lifted for $\sigma_3\neq0$. Then $m_{\mrm{p},33}^2$ mixes with $m_{\mrm{p},00}^2$ and $m_{\mrm{p},88}^2$ since $m_{\mrm{p},03}^2$ and $m_{\mrm{p},38}^2$ become non-zero as is $m_{\mrm{p},08}^2$.
%The mass of the $\pi^0$-meson results %then 
%from diagonalising the $0-3-8$ sector \new{\cite{Chatterjee:2011jd}}.
Using the in-medium pion masses \eq{eq:MedMesMas} the pressure of the pions is strictly speaking a field-dependent correction to the thermodynamical potential that contributes as well to the equations of motion, $\pd\Omega/\pd\varphi_i-\pd{p_\pi}/\pd\varphi_i$ and the (pseudo)scalar masses themselves. In a lowest-order approximation this correction is neglected, saying that the dynamics of the system is governed by the grand canonical potential $\Omega$ alone which determines the meson masses as well. An uncoupled pion gas with these pion masses is then added to thermodynamics as in \Eq{eq:ptot}. We will show how the thermal fluctuations of mesons alter the results for the order parameters, for the meson masses themselves and for thermodynamics in \Ref{Zacchi:2016}.\\
The pion contribution is the dominant one compared to the heavier mesons in the phase where chiral symmetry is broken as it is also shown in \Refs{Herbst:2013ufa,Torres-Rincon:2016ahl}. Furthermore, the more massive mesons would have a sizable contribution in the high temperature phase \cite{Torres-Rincon:2016ahl} where, physically, the mesonic degrees of freedom dissolve. But this aspect of deconfinement is not covered by centre symmetry restoration. This justifies only considering the contribution of a gas of thermal pions but no other mesons.\\

The inclusion of quantum fluctuations of mesons in a renormalisation group framework would not further improve the agreement for the order parameters and thermodynamics within the uncertainty of the parameters  as is shown in \Ref{Herbst:2013ufa}.

%s%s%s%s%s%s%s%
\section{Phase Structure at non-zero Quark Chemical Potentials}\label{sec:ResultsMuQ}
%s%s%s%s%s%s%s%

In the previous section the unquenching of the Polyakov-loop potential showed to be an important improvement for the description of the phase structure and thermodynamics at zero quark chemical potentials with Polyakov-loop--extended chiral models.
Now, the impact of unquenching the Polyakov-loop potential on the phase structure at non-vanishing net quark density will be discussed.

The results presented in the following are obtained using the polynomial-logarithmic Polyakov-loop potential.

\begin{figure}%[t]
	\centering
	\includegraphics[width=.49\textwidth]{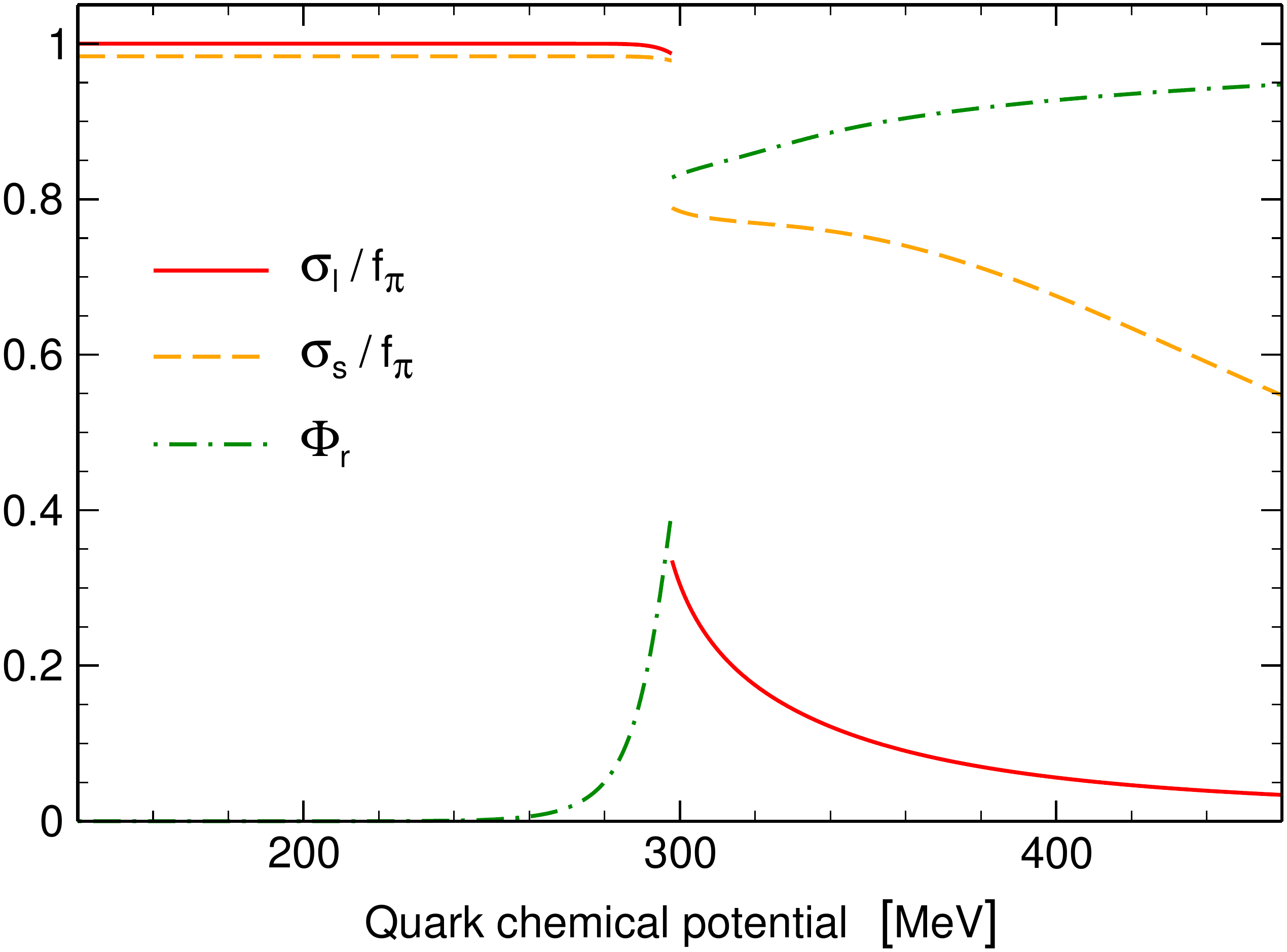}
	\hfill
	\includegraphics[width=.49\textwidth]{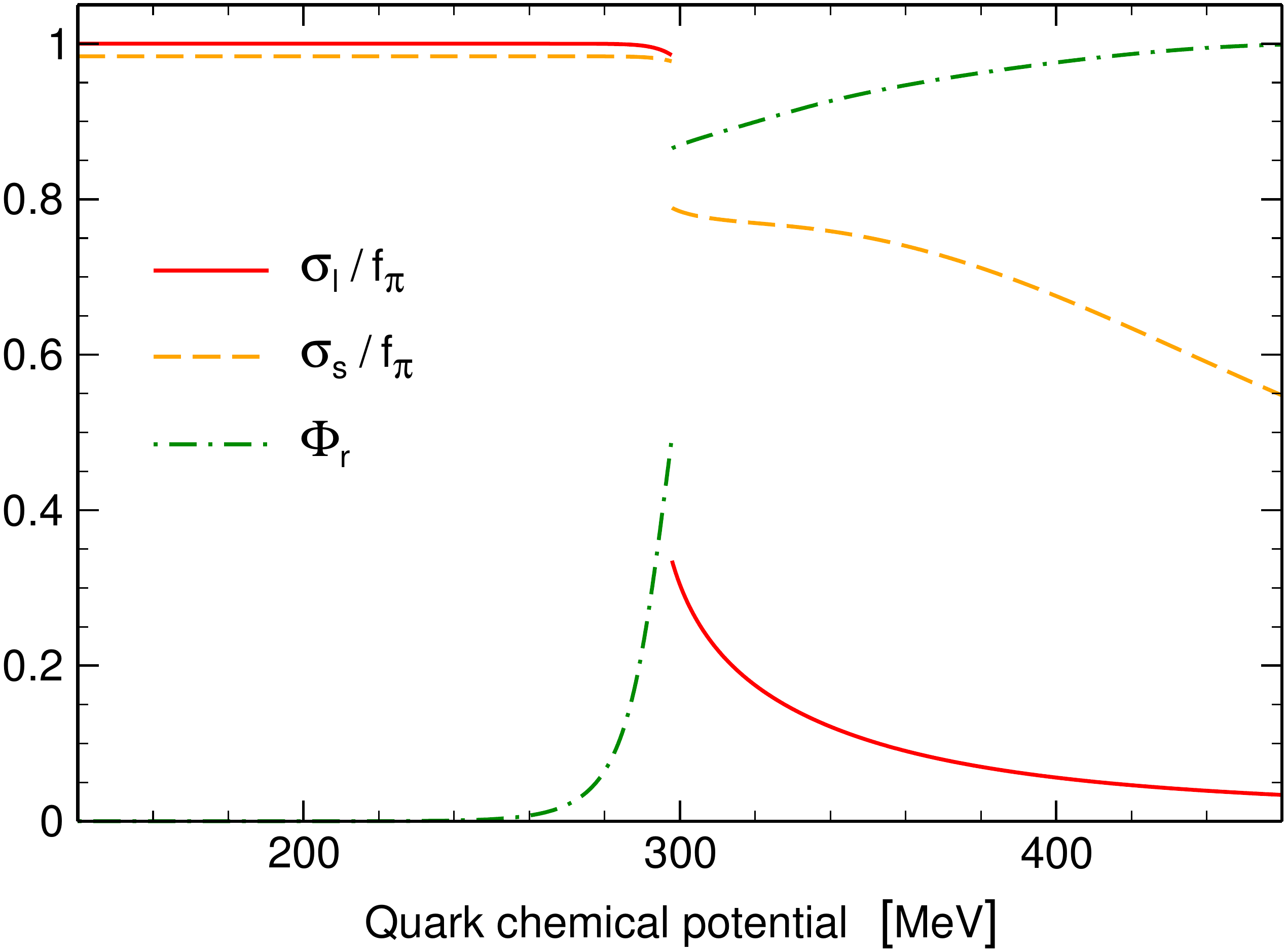}
	\vskip2ex
	\includegraphics[width=.49\textwidth]{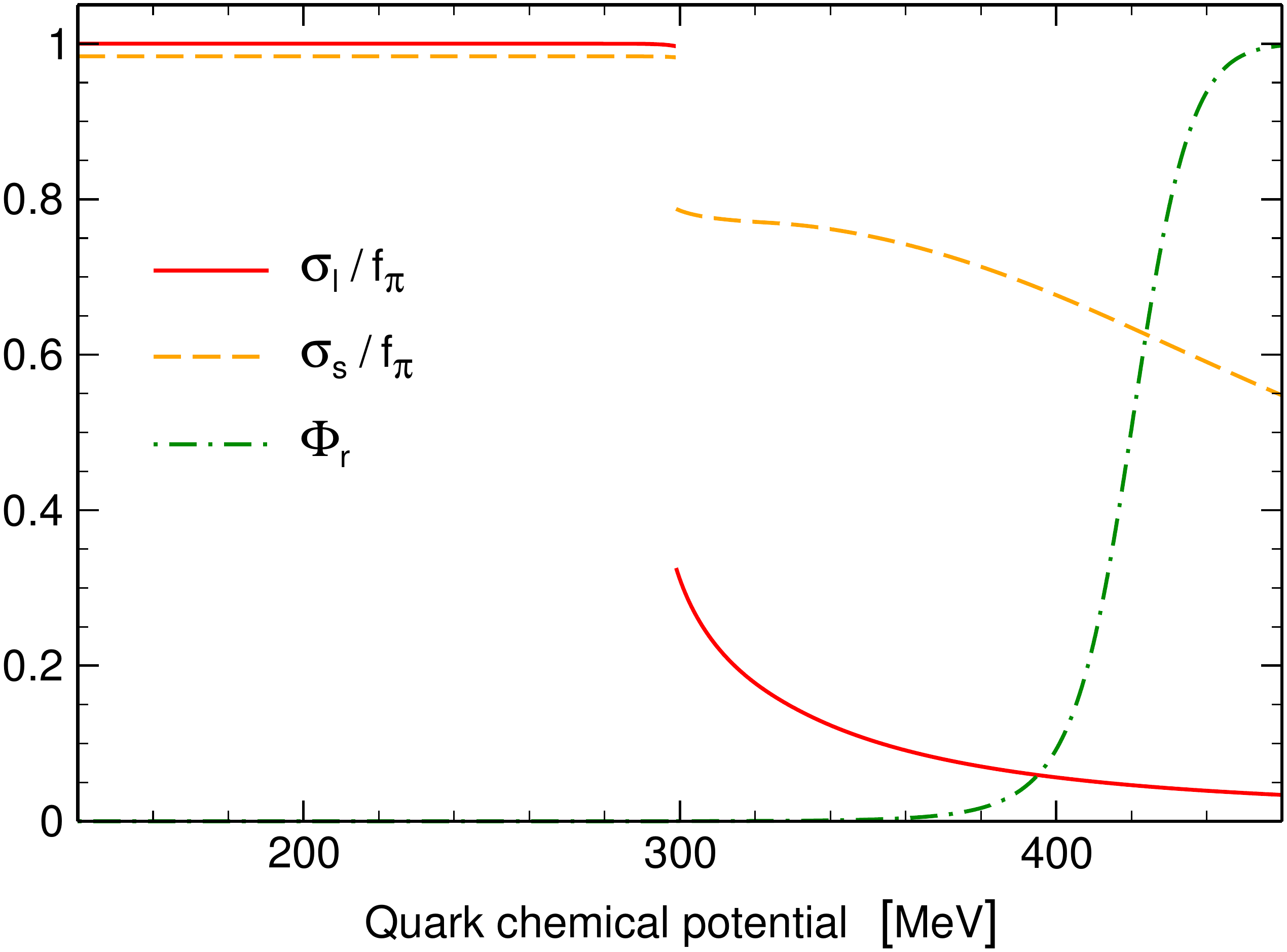}
	\hfill
	\includegraphics[width=.49\textwidth]{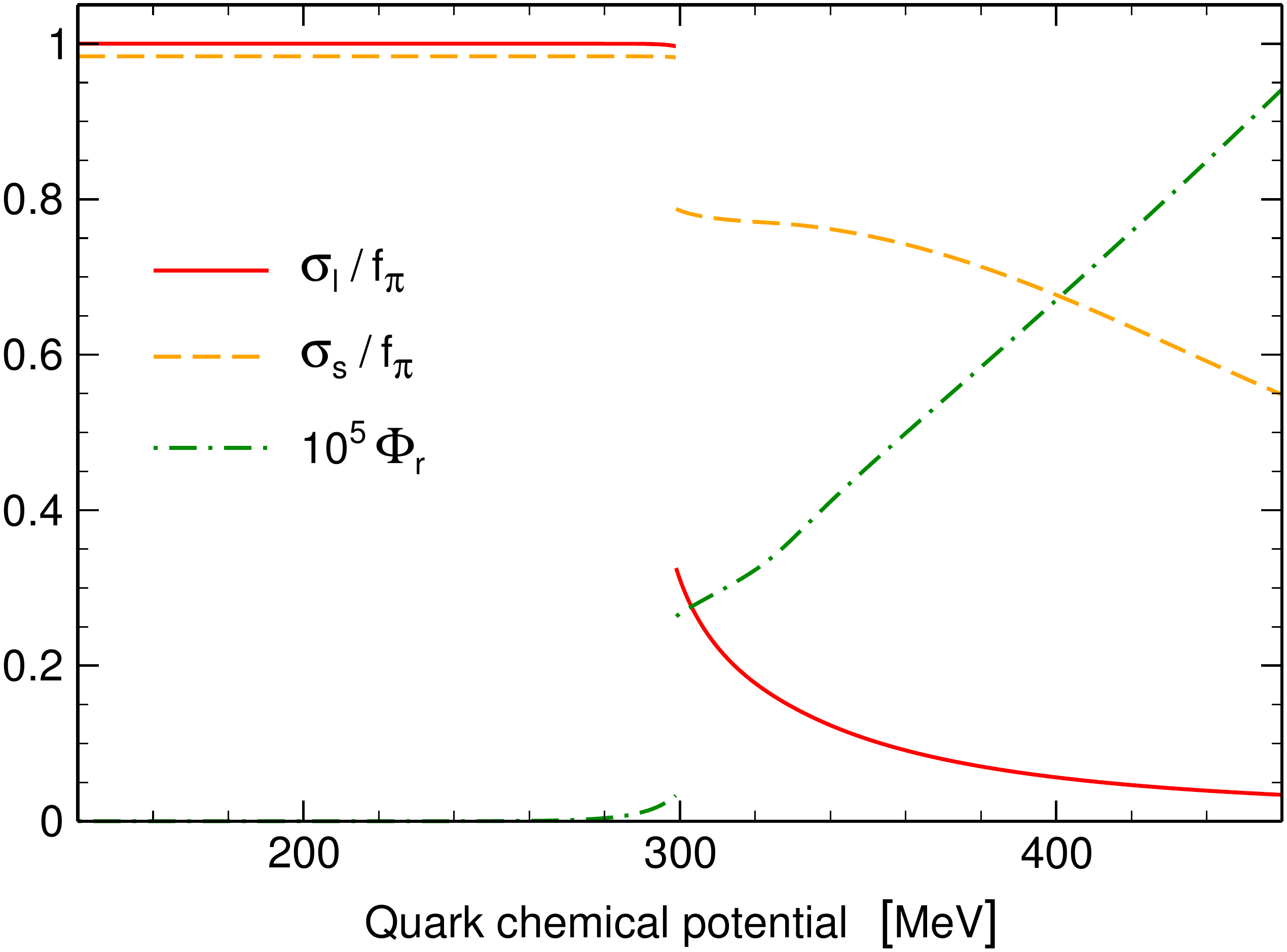}
	\caption{Evolution of the chiral condensates and the Polyakov loop with increasing quark chemical potential at a temperature of 10\,MeV with the unquenched Polyakov-loop potential with a density independent $T_\mrm{cr}^\mrm{glue}$ (top left),  with $T_\mrm{cr}^\mrm{glue}\blr{\mu_f}$ (top right), with the Yang-Mills Polyakov-loop potential with $T_\mrm{cr}^\mrm{YM}\blr{\mu_f}$ (bottom left) and with a density independent $T_\mrm{cr}^\mrm{YM}$ (bottom right).}
	\label{fig:orderparsT10_YMglue}
\end{figure}

The impact of a quark backreaction on the Polyakov-loop potential at finite quark densities is shown in \Fig{fig:orderparsT10_YMglue}. The upper panels are results obtained with the unquenched Polyakov-loop potential while for the results in the lower part a pure Yang-Mills Polyakov-loop potential has been used.
%f%f%f%
%f%f%f%
Obviously, using the quark-enhanced Polyakov-loop potential the chiral and (de)confinement transitions remain linked at small temperatures and large quark chemical potentials.\\
The question is whether the use of the same functional relation between pure Yang-Mills and the glue effective potential found in the transition region for zero chemical potentials is justified at small temperatures and large quark chemical potentials. At small temperatures and zero density one should actually expect that Yang-Mills and the effective glue potential become asymptotically the same. So the question specifies to whether the impact of quarks to the Polyakov-loop potential at small temperatures and chemical potentials of the order of the critical one can be approximated  by their impact at zero density and temperatures around the pseudocritical temperature.\\
The pure Yang-Mills Polyakov-loop potential as such does not contain any explicit dependence on quark densities and therefore, the Polyakov loop remains at its vacuum value $\sim\!0$ at small temperatures when the quark chemical potential is increased as is shown in the lower right part of \Fig{fig:orderparsT10_YMglue}. Only the coupling to quarks and mesons leads to an extremely small variation.\\
One way to include a density-dependent matter backreaction in the Polyakov-loop potential is to consider the chemical potential dependence of its transition temperature \cite{Schaefer:2007pw,Herbst:2013ail}. This implies that the transition scale of the Polyakov-loop potential decreases with increasing chemical potential. This in turn lowers the pseudocritical temperature of the crossover transition of the Polyakov loop at a large chemical potential towards the critical temperature of the chiral first-order transition. This is shown in the lower left panel of \Fig{fig:orderparsT10_YMglue} and \Refs{Schaefer:2007pw,Herbst:2010rf,Schaefer:2011ex,Herbst:2013ail}. References \cite{Herbst:2010rf,Herbst:2013ail}  included meson fluctuations in a renormalisation group framework and they observed that this backreaction is then enough to see the chiral and (de)confinement transitions remaining linked at large chemical potentials.\\
Therefore, the coincidence of the chiral and (de)confinement transitions at large chemical potentials with the unquenched Polyakov-loop potential is qualitatively a welcome feature. Taking additionally into account the chemical potential dependence of the glue critical temperature shows then a minor impact as can be seen when comparing the upper right and upper left figures of \Fig{fig:orderparsT10_YMglue}. The consequence of adjusting $T_\mrm{cr}^\mrm{glue}$ with $\mu_\mrm{q}$ is as for the Yang-Mills Polyakov-loop potential: it moves the Polyakov loop to larger values at a given chemical potential.\\
Interestingly enough, the evolution of the chiral order parameters is perfectly independent of the realisation of the Polyakov-loop potential as is shown in \Fig{fig:orderparsT10_YMglue}, except for a $1\,\%$ decrease of the chiral condensates just before the transition when the unquenched Polyakov-loop potential is used.

Even though the qualitative impact of unquenching the Polyakov-loop potential (that it links the chiral and (de)confinement transitions) is reasonable its quantitative magnitude remains in question.
While pure gauge theory allows to adjust parametrisations of the Polyakov-loop potential to the minimum of the potential and the PQM/PNJL model at zero densities but non-zero temperature probes regions of the Polyakov-loop potential away from the minimum, the PQM/PNJL model at large quark chemical potentials probes the form of the Polyakov-loop potential far away from the minimum.
The impact of unquenching the Polyakov-loop potential is such that the Polyakov loop becomes unbound at small temperatures and large chemical potentials for polynomial parametrisations. Only the limitation of the Polyakov loop to be smaller than one when the Haar measure is considered avoids this behaviour.\\

\begin{figure}%[t]
	\centering
	\includegraphics[width=.49\textwidth]{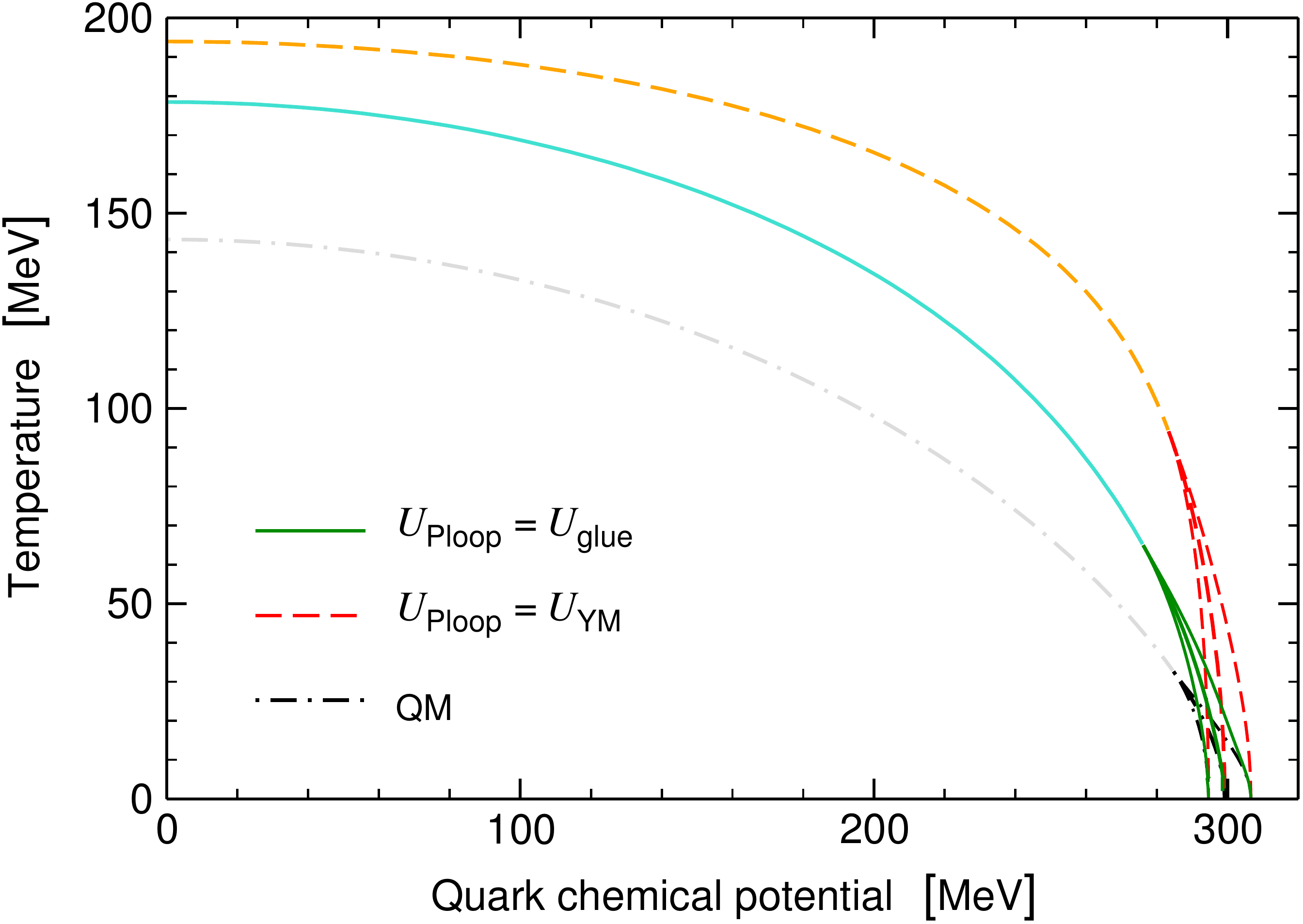}
	\hfill
	\includegraphics[width=.49\textwidth]{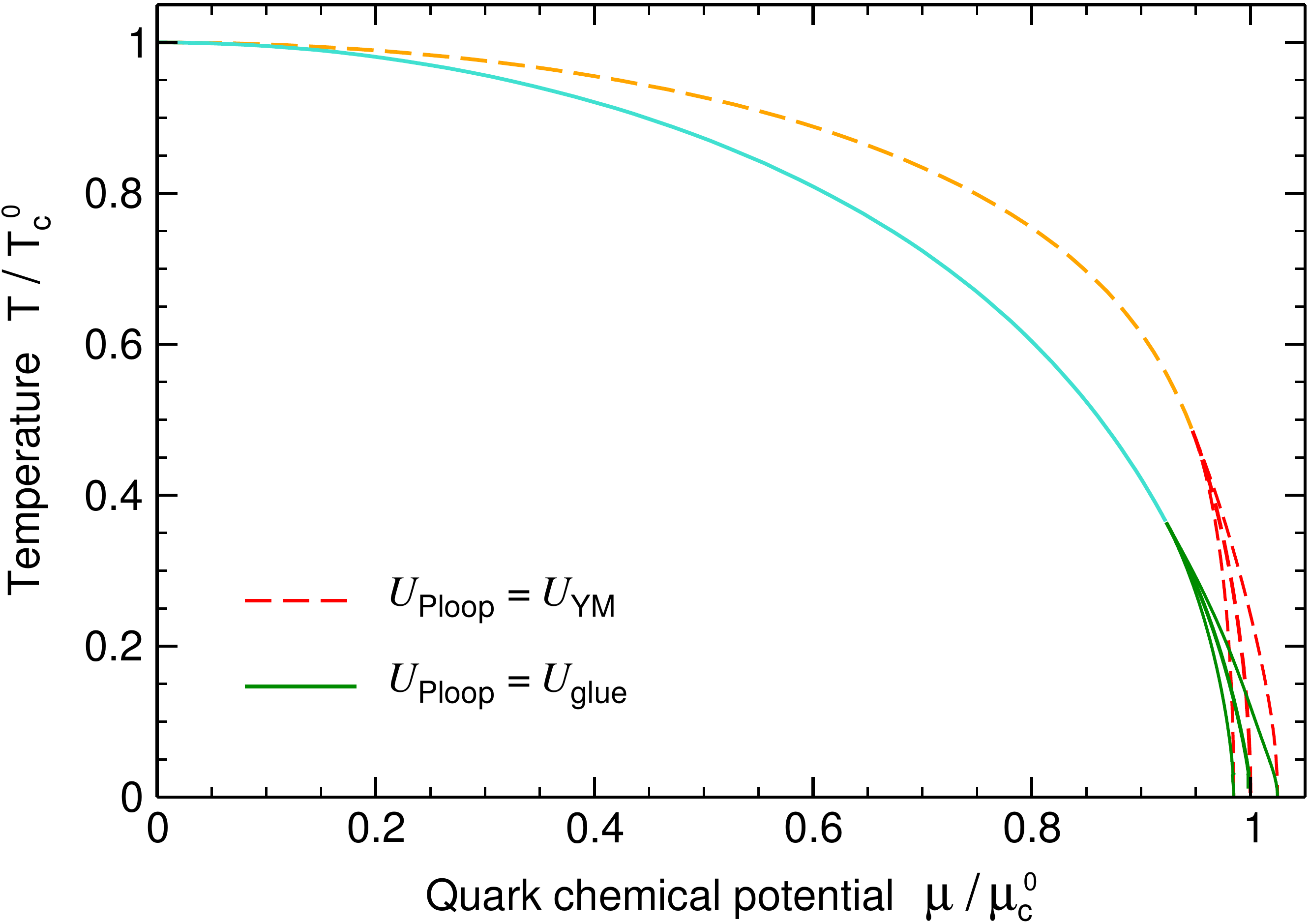}
	\caption{Phase diagram with the Yang-Mills Polyakov-loop potential and the unquenched one, in absolute units (left) and relative units (right). The (lighter) crossover lines are the pseudocritical values of the subtracted chiral condensate. The thinner, outer lines around the first-order transition lines retrace the extension of the metastable regions.}
	\label{fig:PhaseDiagram_YMglue}
\end{figure}

Having found a crossover at small chemical potentials and a first-order transition at small temperatures, the next step is to investigate the complete $T-\mu_\mrm{q}$ phase diagram considering the above-mentioned observations.\\
The phase diagram of the PQM model is shown in \Fig{fig:PhaseDiagram_YMglue}. The effect of unquenching the Polyakov-loop potential on the phase diagram is displayed by the comparison of the results using the Yang-Mills Polyakov-loop potential and its quark-enhanced counterpart. The (lighter) crossover lines are the pseudocritical values of the subtracted chiral condensate.\\
%f%f%f%
%f%f%f%
For all values of the quark chemical potential $\mu_\mrm{q}=\mu_\mrm{u}=\mu_\mrm{d}=\mu_\mrm{s}$, unquenching the Polyakov-loop potential has the consequence of lowering the transition temperature which has its origin in the effect of lowering the transition scale of the Polyakov loop.
In the zero-temperature limit $T\to0$ the Polyakov-loop potential becomes independent of its variables $\mc{U}\blr{\Phir,\Phii;T=0}=0$ $\forall\blr{\Phir,\Phii}$ since gluon excitations are entirely independent of the quark chemical potential and therefore, the phase structure is that of the Quark-Meson model that is shown as well in \Fig{fig:PhaseDiagram_YMglue}.\\
With all  uncertainties adjusted to lattice data at zero quark chemical potential, there remains a region at large chemical potentials where the phase transition is discontinuous and the order parameters show a jump. The thinner, outer lines around the first-order transition line retrace the extension of the metastable region, i.e., to which extent the one phase remains as a local minimum while the system is in the other phase which is the global minimum. The coordinates of the critical endpoints (CEP) are $\blr{T,\mu}_\mrm{CEP}=\blr{65,276}\,\mrm{MeV}$ with the unquenched Polyakov-loop potential and $\blr{94,283}\,\mrm{MeV}$ with the Yang-Mills potential. So the main effect of including the quark backreaction on the gluons is to lower the temperature of the CEP at a similar chemical potential.
Interestingly enough, \Ref{Biguet:2014sga} did a detailed analysis of the quantitative sensitivity of the CEP coordinate to the meson phenomenology in the vacuum that is used to adjust the parameters in the chiral sector. They also find a much larger sensitivity of $T_\mrm{CEP}$ than $\mu_\mrm{CEP}$, this time for variations of the chiral properties.

At these large values of the quark or baryon chemical potential ($\mu_\mrm{b}=3\mu_\mrm{q}$), baryons are the relevant degrees of freedom that are not contained in the PQM model. To implement baryons as bound states of quarks and diquarks using the Fadeev and Bethe-Salpeter equations, see e.g.~\Refs{Eichmann:2010je,Oettel:2000ig,Ebert:1997hr,Buck:1992wz}, to the PQM model remains for future work.  For work on including diquarks in two-colour QCD, see e.g.~\Refs{Strodthoff:2011tz,Strodthoff:2013cua}. Reference \cite{Eichmann:2015kfa} included baryonic degrees of freedom for their localisation of the CEP but finds only a small influence of these within a few MeV.
Even though this region of the phase diagram at large chemical potentials and small temperatures is speculative within the present framework, general methods to analyse the properties of this region can be discussed, as in \Sec{sec:Nucleation}.

An important aspect of the phase diagram at small chemical potentials is the curvature of the transition line. This can also be extracted from lattice calculations that are hampered at non-zero quark chemical potentials due to the sign problem. In the right part of \Fig{fig:PhaseDiagram_YMglue} the transition lines are normalised by the respective pseudocritical temperatures at zero density, $T_c^0$ and the critical chemical potential at zero temperature, $\mu_\mrm{c}^0$ to allow for a comparison of the curvatures. One sees that applying the unquenched Polyakov-loop potential lowers the pseudocritical temperature relatively stronger with increasing chemical potential such that the curvature of the phase transition line increases.

\begin{figure}%[t]
	\centering
	\includegraphics[width=.49\textwidth]{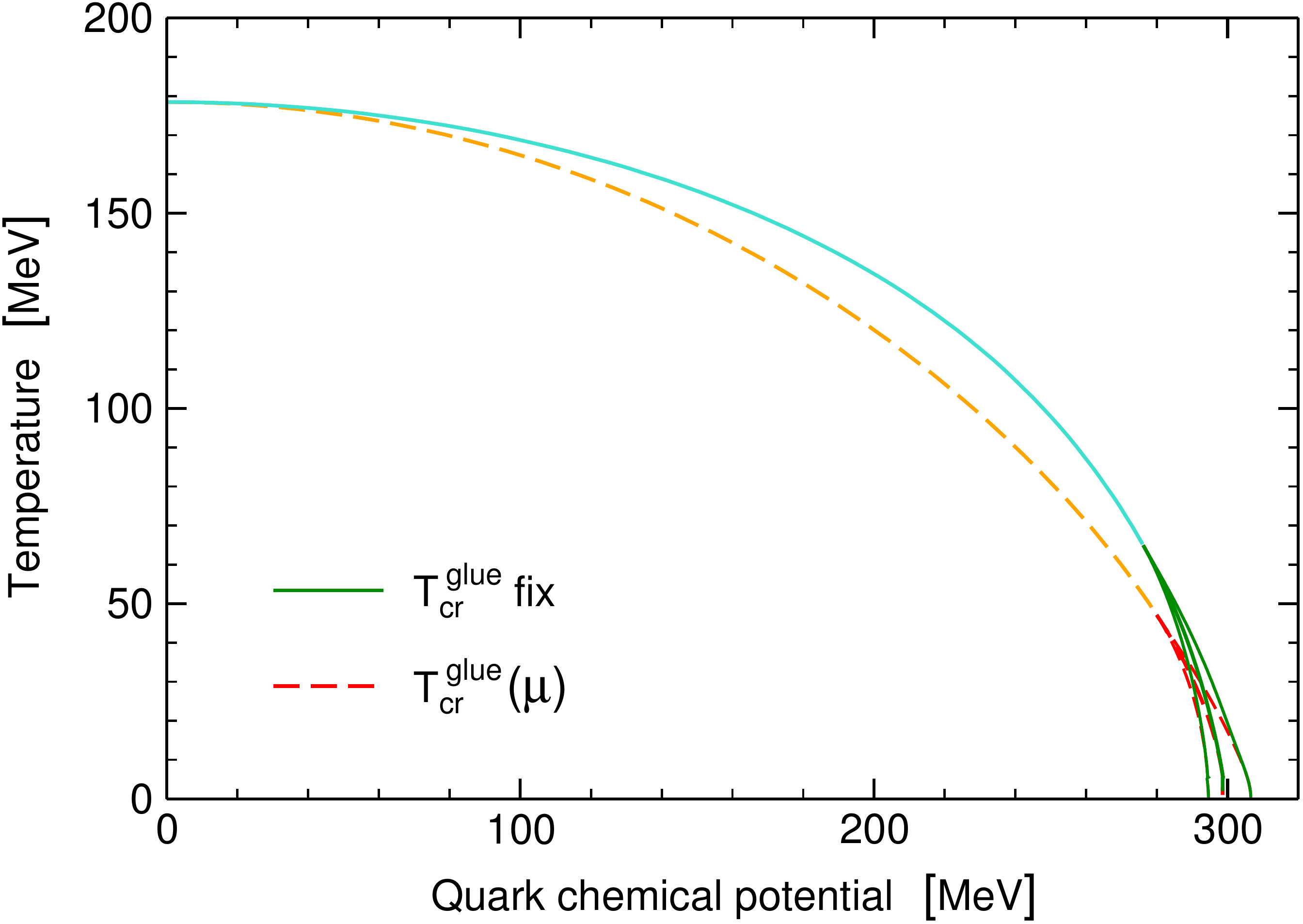}
	\hfill
	\includegraphics[width=.49\textwidth]{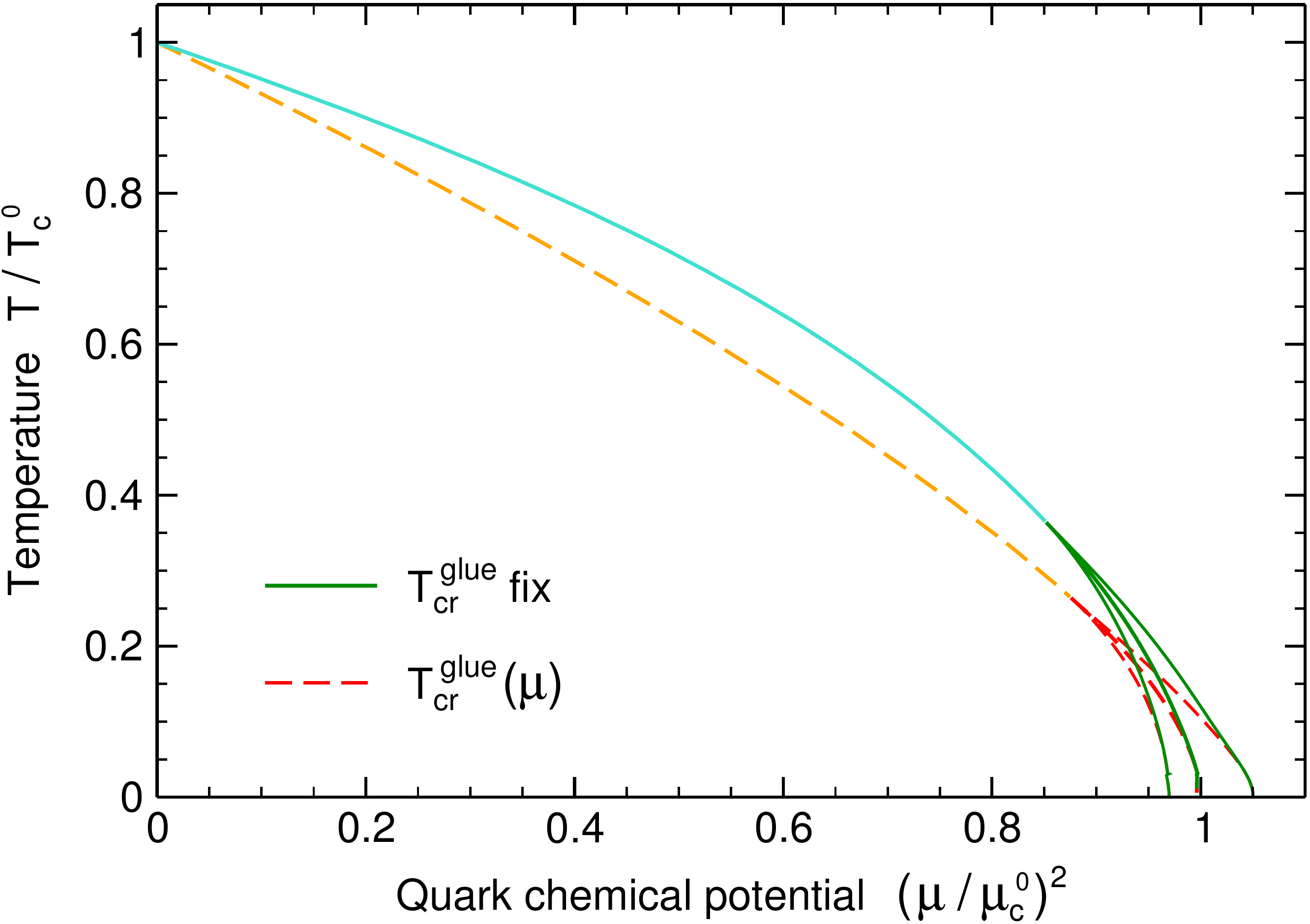}
	\caption{Phase diagram for a constant critical scale of the unquenched Polyakov-loop potential and with the density dependence of \Ref{Schaefer:2007pw}, in absolute units (left) and relative units (right).}
	\label{fig:PhaseDiagram_T0fixmu}
\end{figure}

The influence of lowering the critical temperature of the Polyakov-loop potential with increasing density has a smaller effect on the phase diagram than unquenching the Polyakov-loop potential, as can can be seen in the comparison of Figs.~\ref{fig:PhaseDiagram_YMglue} and \ref{fig:PhaseDiagram_T0fixmu}.
%f%f%f%
%f%f%f%
To illustrate the impact on the curvature of the transition line, in the right part of \Fig{fig:PhaseDiagram_T0fixmu} the square of the chemical potential is chosen as the abscissa. In this unit the pseudocritical temperature follows a straight line for small values of the chemical potential. Lowering the glue critical temperature with increasing chemical potential has the impact of decreasing the pseudocritical temperature such that the curvature of the transition line gets larger.
The location of the critical endpoint is lowered further in temperature to $\blr{T,\mu}_\mrm{CEP}=\blr{47,280}\,\mrm{MeV}$ when the glue critical temperature is decreased with growing chemical potential.

\begin{figure}%[h]
	\includegraphics[width=0.6\textwidth]{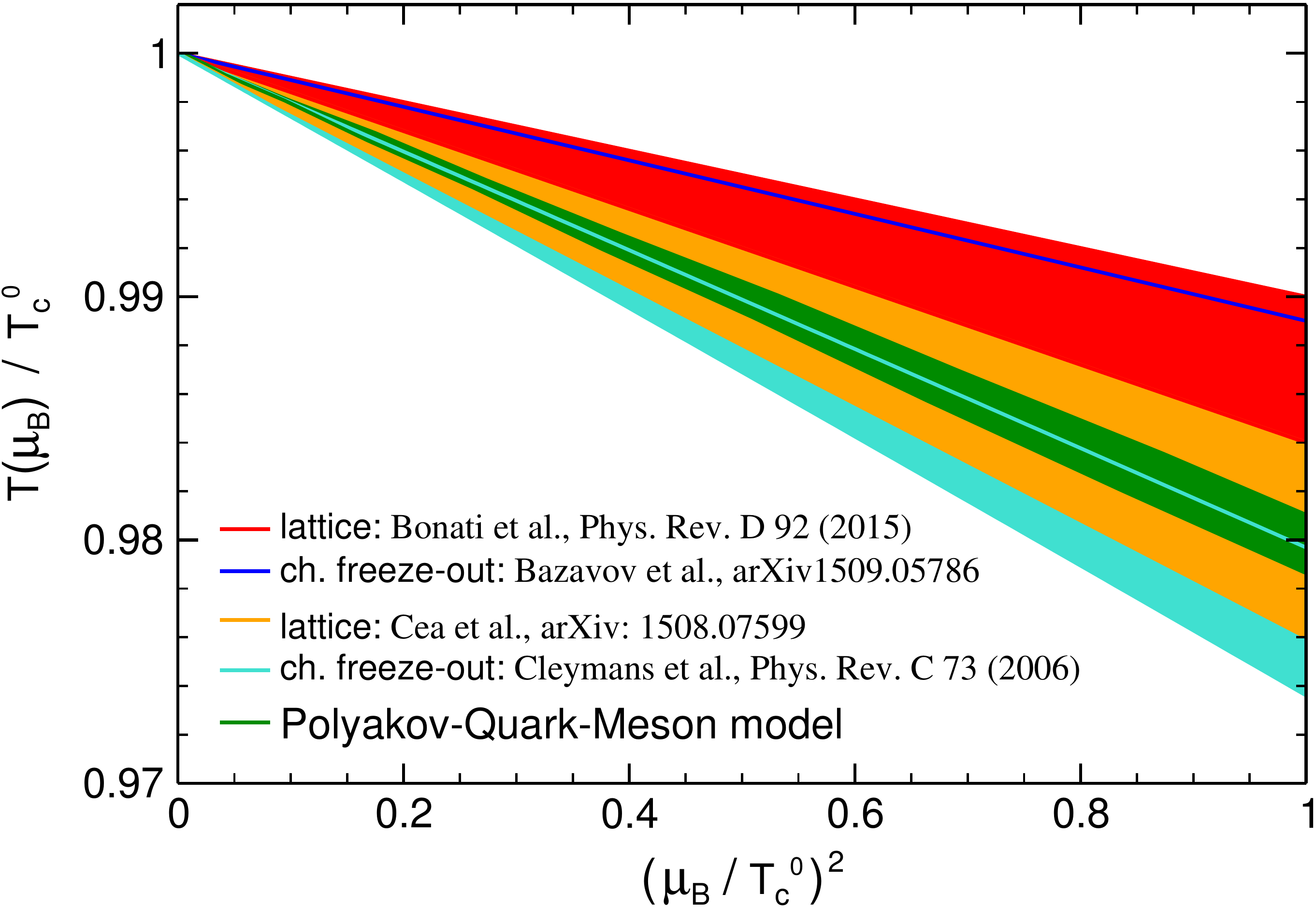}
	\hskip-10ex
	\begin{minipage}[b]{0.5\textwidth}
		\caption{\label{fig:phdiag_curvature} Curvature of the pseudocritical phase transition line. Our result is compared to those of recent lattice calculations \cite{Bonati:2015bha,Cea:2015cya} and the location of the latest hadron chemical equilibrium in relativistic heavy-ion collisions \cite{Cleymans:2005xv,Bazavov:2015zja}. The result of \Ref{Bazavov:2015zja} is a lower bound.}
	\end{minipage}
\end{figure}

The curvature of the transition line is also a quantity that can be extracted from lattice calculations up to baryon chemical potentials for which terms of higher order than $(\mu_\mrm{B}/T_\mrm{c})^2$ can be neglected. Therefore, it can serve as a test of the predictive power of the presented framework that is complementary to the one that we performed in \Ref{Stiele:2013pma} for the isospin dependence of the pseudocritical temperature. Figure \ref{fig:phdiag_curvature} compares the result using the unquenched Polyakov-loop potential with density-independent glue critical temperature with those of the recent lattice calculations of \Refs{Bonati:2015bha,Cea:2015cya}.
%f%f%f%
%f%f%f%
We show these since they span the range of results of the latest generation of lattice calculations. The one of \Ref{Bellwied:2015rza} is as well within this range. The result of the presented framework with its uncertainties adjusted at zero density is well within the ballpark of these lattice results within their uncertainties. The band shown for the PQM model results from the difference between using the logarithmic or polynomial-logarithmic Polyakov-loop potential. This difference is further analysed in \Sec{su:Pardep}.
Also shown in \Fig{fig:phdiag_curvature} are the curves on which chemical freeze-out of hadrons is expected according to the beam-energy dependence of observed particle yields in relativistic heavy-ion collisions  \cite{Cleymans:2005xv,Bazavov:2015zja}. Compared to the result in \Ref{Cleymans:2005xv} the outcome of our calculation fulfils the expectation that the transition line lies close above the freeze-out curve. A comparison to the  lower limit given in \Ref{Bazavov:2015zja} would violate this expectation, an observation which also holds true for most of the lattice results within their errors.\\
One ingredient to the presented model that diminishes the curvature of the crossover line is the inclusion of meson fluctuations in a renormalization group framework \cite{Herbst:2010rf,Herbst:2013ail}.
In \Ref{Bratovic:2012qs} it was quantified how including repulsive vector interactions reduces the slope of the phase transition line. However, %But 
one should be aware that when including the vector-meson exchange, the model still fails to describe lattice results of quark number susceptibilities \cite{Steinheimer:2010sp,Steinheimer:2014kka}, see further the discussion in \Ref{Restrepo:2014fna}.

\begin{figure}
	\centering
	\includegraphics[width=.49\textwidth]{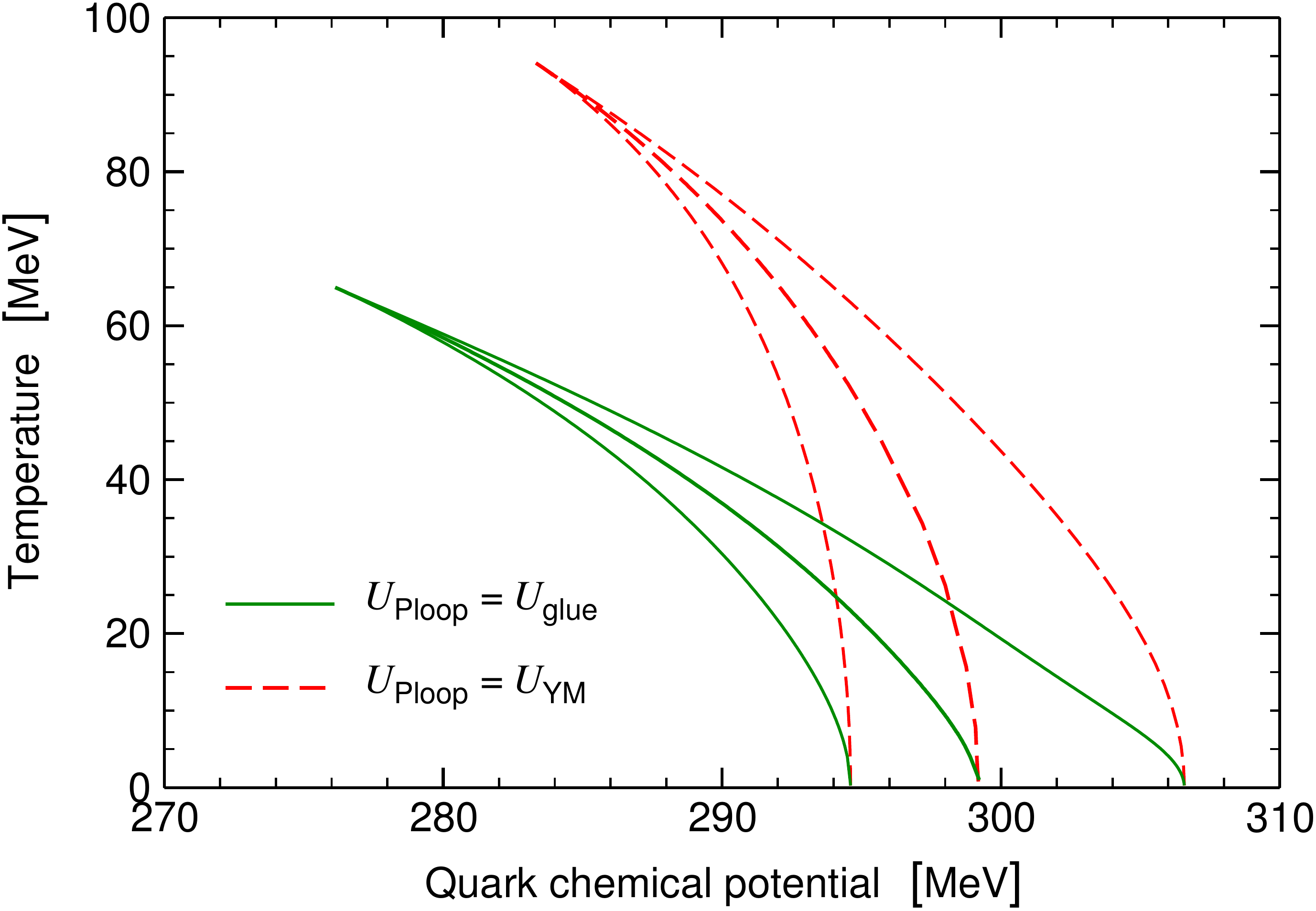}
	\hfill
	\includegraphics[width=.49\textwidth]{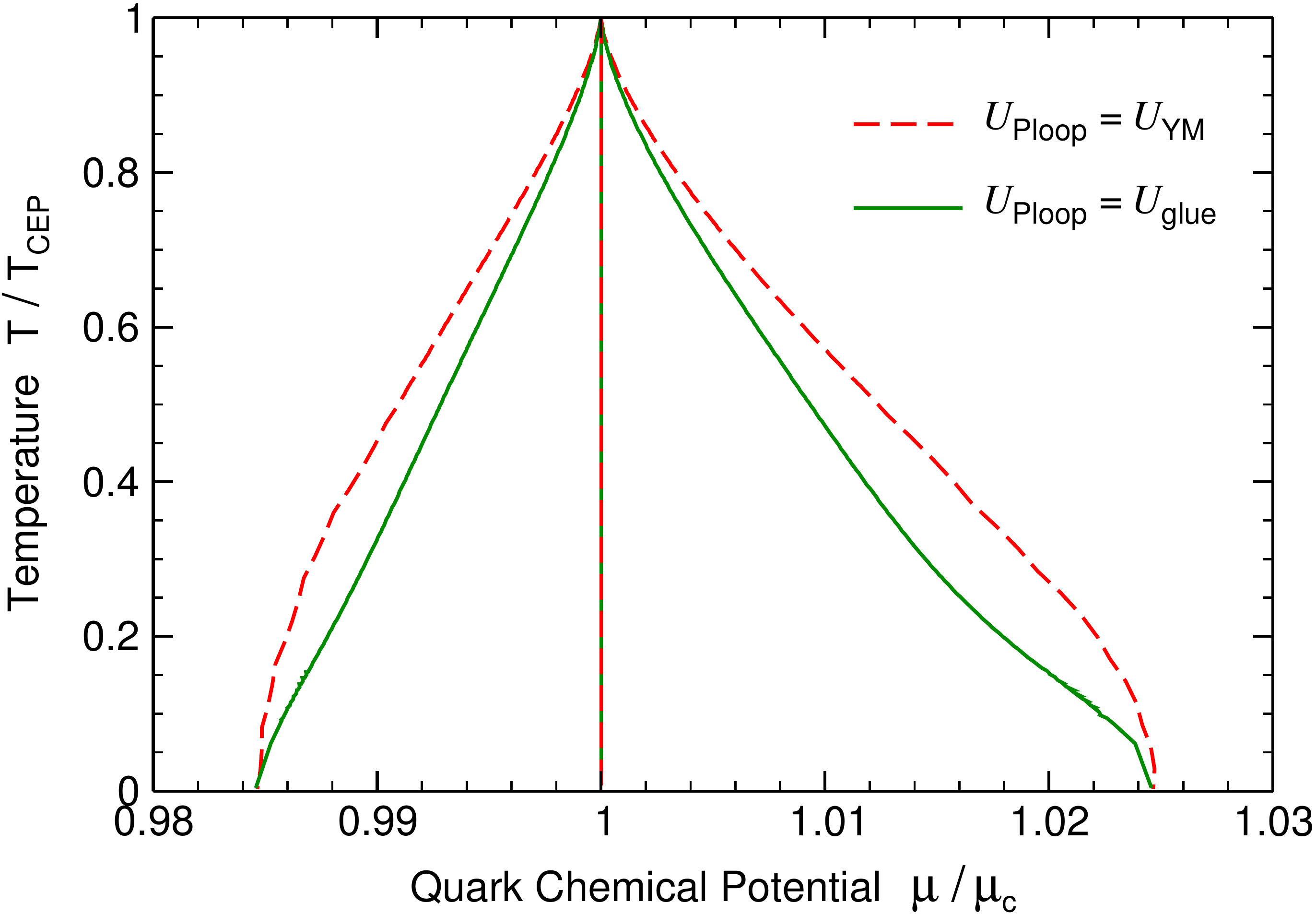}
	\caption{Metastable regions of the PQM model with the Yang-Mills Polyakov-loop potential and the unquenched one. Their extension is compared in absolute units (left) and with respect to the transition line ($\mu_\mrm{c}$) and the temperature of the critical endpoint $T_\mrm{CEP}$ (right).}
	\label{fig:MetastableRegion_YMglue}
\end{figure}

In general, the two minima of the potential that are degenerate on the coexistence line persist as a global and a metastable local minimum in some region of the phase diagram around the phase transition line. Going away from the coexistence line, the intervening maximum approaches the local minimum until these two extrema meet and form an inflection point that defines the spinodal line.\\
Figure \ref{fig:MetastableRegion_YMglue} compares the extension of the metastable region that is limited by the spinodal lines using the unquenched Polyakov-loop potential and the Yang-Mills potential. 
%f%f%f%
%f%f%f%
The absolute location of the critical endpoint and therefore of the metastable region differs except for the lowest temperatures when the Polyakov-loop independent Quark-Meson limit is reached. But the relative extent is similar as is shown on the right side of \Fig{fig:MetastableRegion_YMglue}. Here, the form changes from a convex shape with the Yang-Mills Polyakov-loop potential to a concave form when the matter backreaction to the Polyakov-loop potential is considered.

%sub%sub%sub%
\subsection{\label{su:Pardep}Dependence on the parametrisation of the Polyakov-loop potential\protect}
%sub%sub%sub%

\begin{figure}%[t]
	\centering
	\includegraphics[width=.49\textwidth]{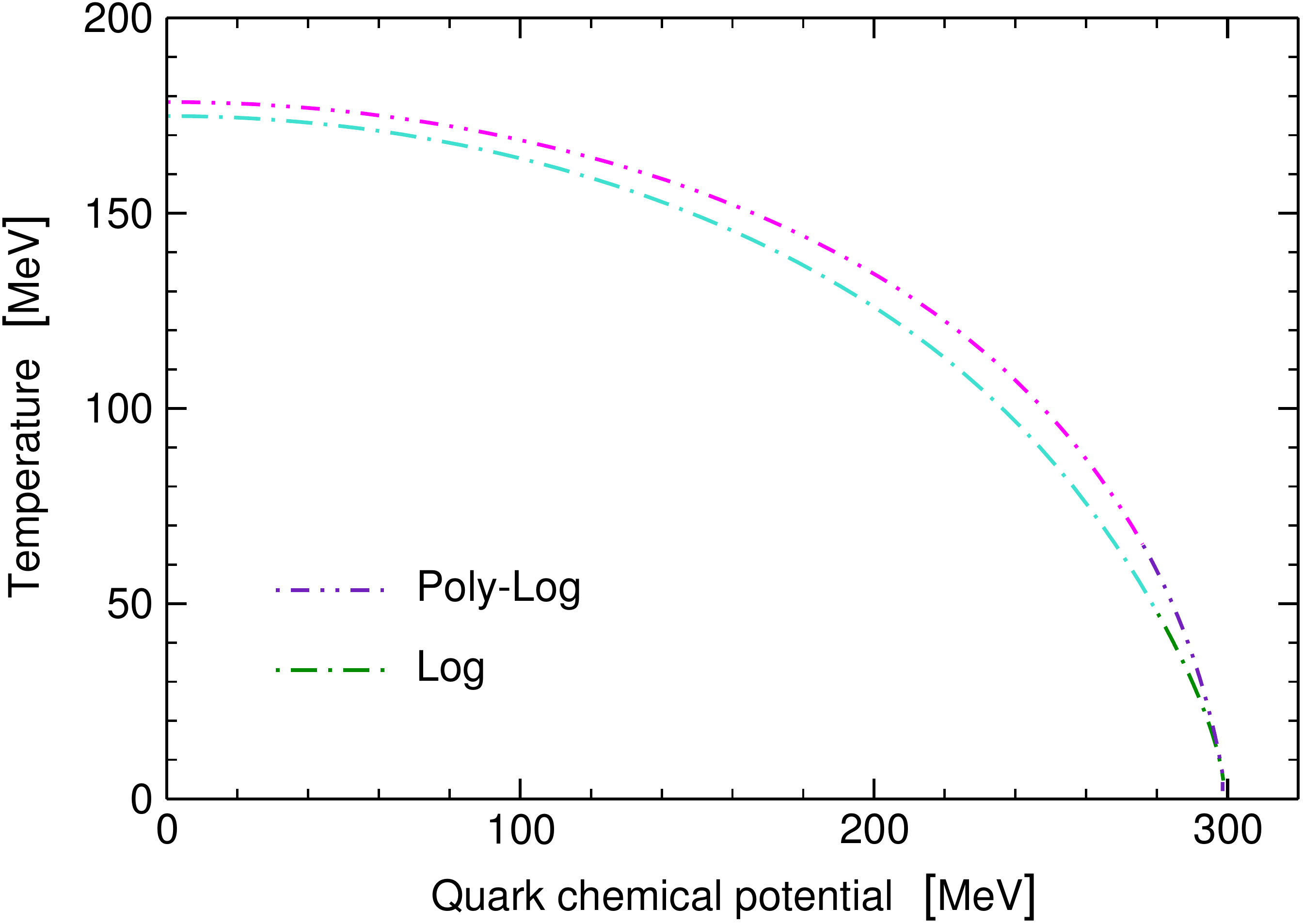}
	\hfill
	\includegraphics[width=.49\textwidth]{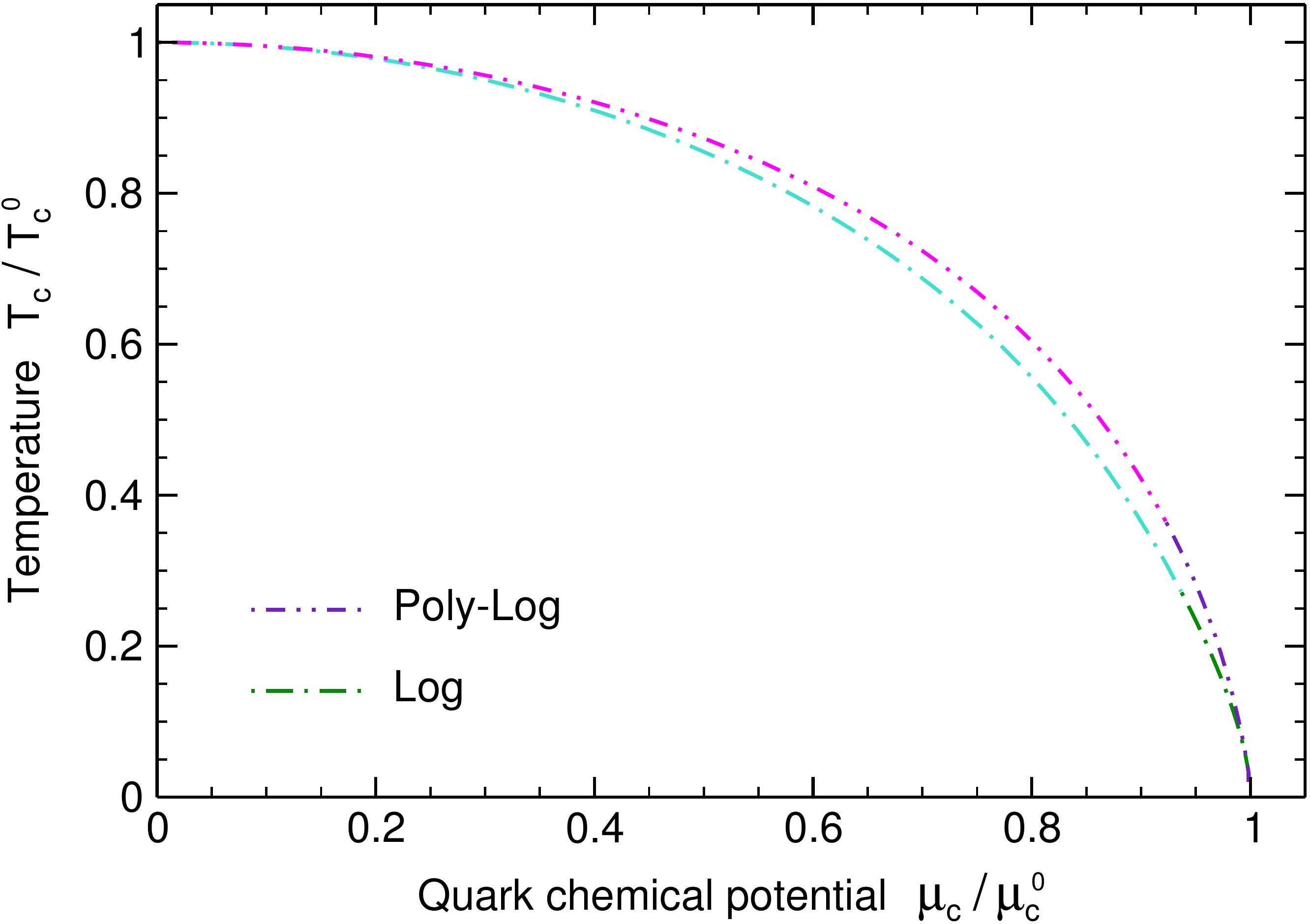}
	\caption{Phase diagram for the different parametrisations of the Polyakov-loop potential, in absolute units (left) and relative units (right).}
	\label{fig:PhaseDiagram_params}
\end{figure}

The analysis of how well the different parametrisations of the Polyakov-loop potential are adjusted to lattice data of Yang-Mills theory shown in \Tab{tab:Ploop_pot_data} and the investigation of the impact of the parametrisations on the evolution of order parameters and thermodynamics at zero chemical potentials when the potential is coupled to quarks and mesons as discussed in \Sec{sec:Resultsmu0}, showed that the results show similar qualitative behaviour.
Figure \ref{fig:PhaseDiagram_params} compares the phase diagrams for the different parametrisations using the parameters as they were determined in \Sec{sec:Resultsmu0} to get as close as possible to the lattice results for the order parameters, thermodynamics and pseudocritical temperatures.
%f%f%f%
%f%f%f%
The overall shape of the phase boundaries is similar and the polynomial-logarithmic Polyakov-loop potential leads to relatively larger transition temperatures from medium chemical potentials on.
The location of the critical endpoint is $\blr{T,\mu}_\mrm{CEP}=\blr{47,280}\,\mrm{MeV}$ with the logarithmic Polyakov-loop potential.

\begin{figure}
	\centering
	\includegraphics[width=.49\textwidth]{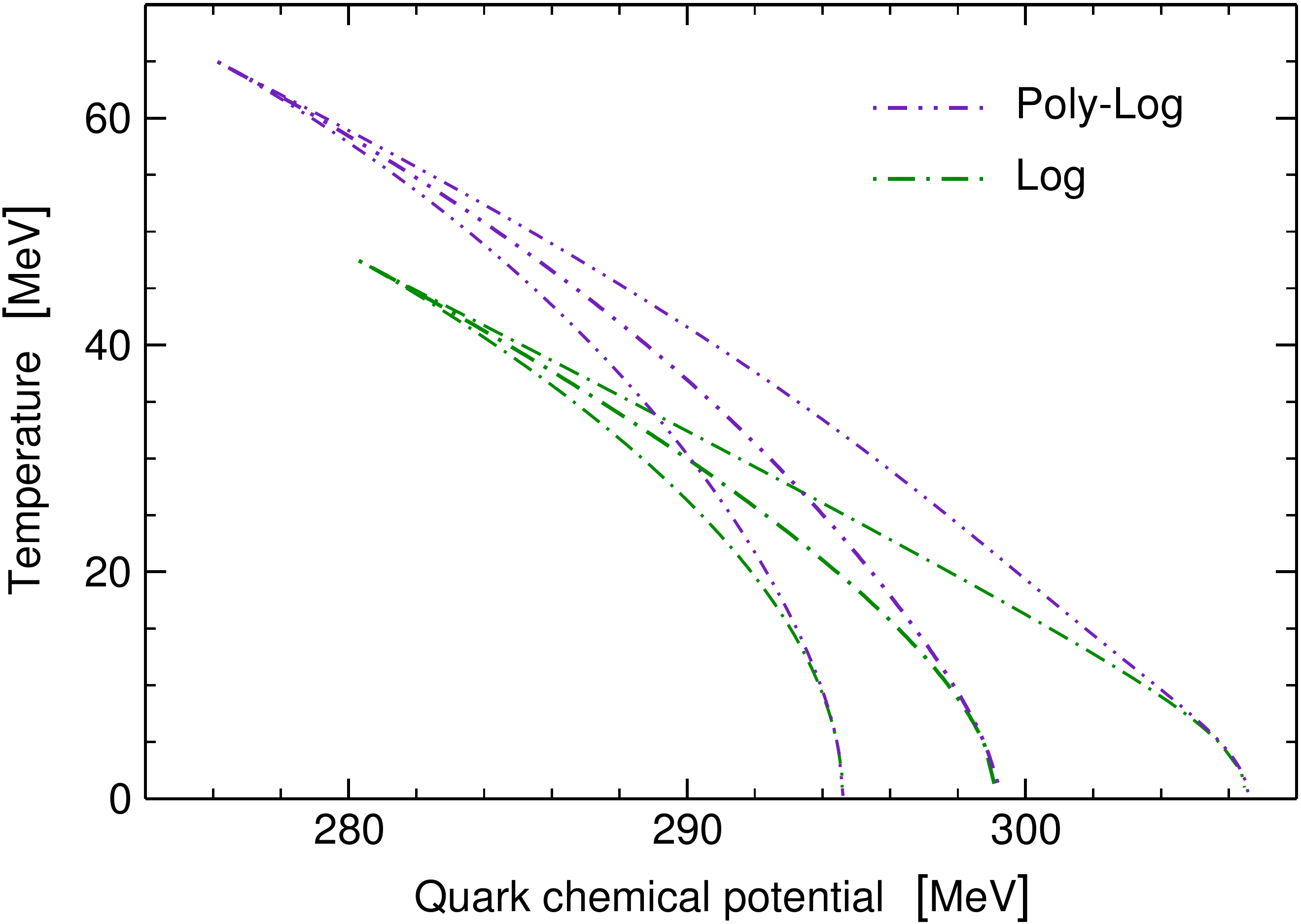}
	\hfill
	\includegraphics[width=.49\textwidth]{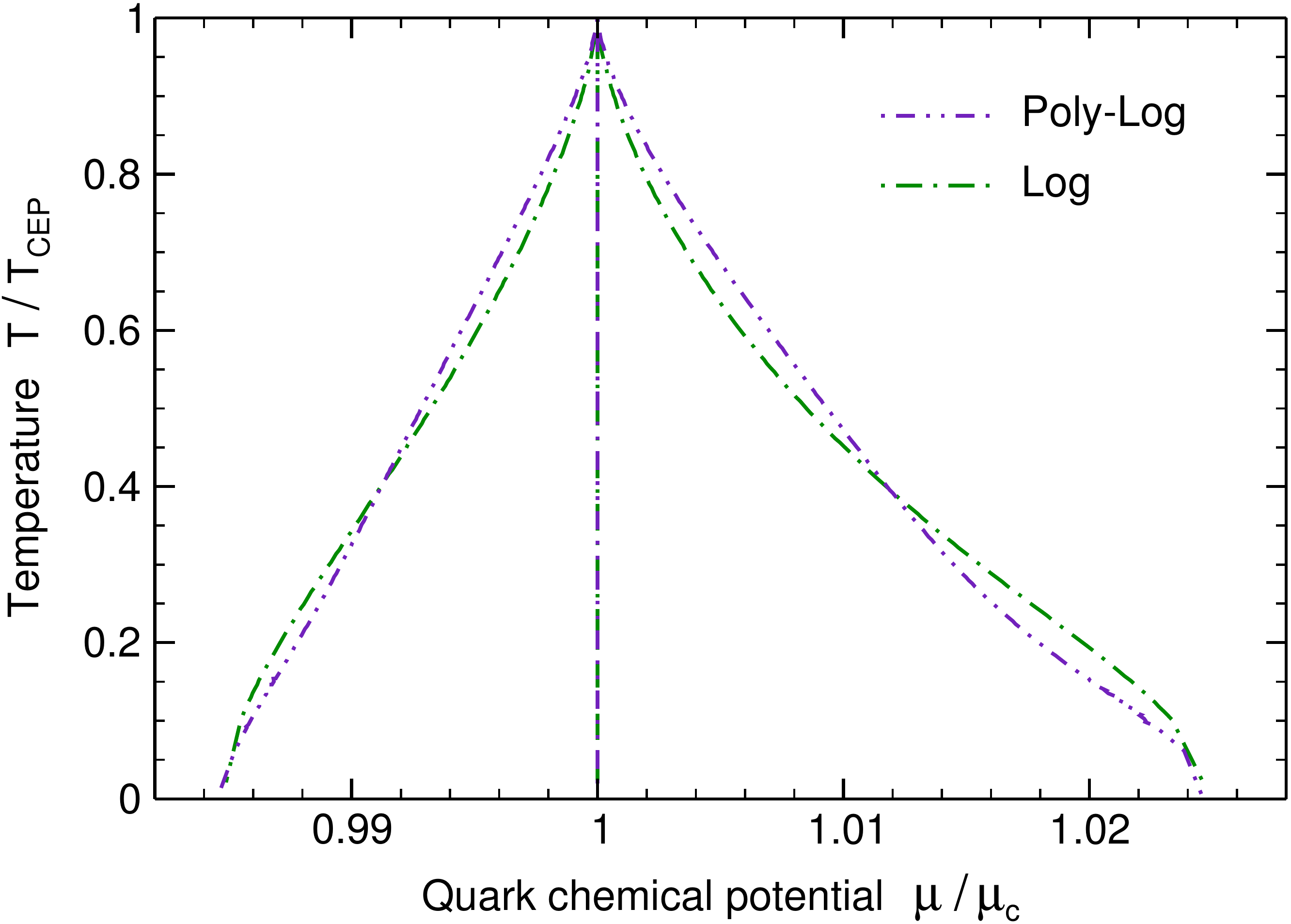}
	\caption{Metastable regions of the PQM model for the different parametrisations of the Polyakov-loop potential. Their extension is compared in absolute units (left) and with respect to the transition line ($\mu_\mrm{c}$) and the temperature of the critical endpoint $T_\mrm{CEP}$ (right).}
	\label{fig:MetastableRegion_params}
\end{figure}

The zoom into metastable regions achieved with different parametrisations of the Polyakov-loop potential displayed in \Fig{fig:MetastableRegion_params} shows in more detail that the critical endpoint with the polynomial-logarithmic glue potential deviates from the location of the critical endpoint calculated with the logarithmic Polyakov-loop potential.
%f%f%f%
%f%f%f%
But once the metastable regions are corrected for the different coordinates of the critical endpoints, both parametrisations lead to an extent of the metastable region that is largely independent of the form of the Polyakov-loop potential. This is  shown on the right side of \Fig{fig:MetastableRegion_params}.
Furthermore, one finds that the degree of metastability that can be reached is relatively modest.
The extent of the metastable region shows an asymmetry for the lowest temperatures in which the spinodal line  in the chirally restored and deconfined phase has a larger distance to the coexistence line than the spinodal in the chirally restored and confined phase.
This observation on the asymmetric extent of the metastable region holds as well for the description of Yang-Mills theory with the different Polyakov-loop potentials and in the pure chiral Quark-Meson model.

%s%s%s%s%s%s%s%
\section{Nucleation in a Polyakov-loop--extended chiral effective model}\label{sec:Nucleation}
%s%s%s%s%s%s%s%

Here, the properties of the phase transition in the high-density and low-temperature region of the phase diagram are analysed.\\
The derivation of an overestimate of the surface tension for bubble nucleation within the thin-wall approximation in \Ref{Mintz:2012aua} shows that it is determined by two contributions. On the one hand by the distance between the two degenerate minima in the space of the order parameters and on the other hand by the shape of the effective potential along the straight line that connects both minima.\\
To discuss how the different contributions to the Polyakov--Quark-Meson model build up the result of the surface tension, \Sec{su:NucleationQM} presents the results within the Quark-Meson model and \Sec{su:NucleationPQM} discusses the effects of the Polyakov-loop extension.

%sub%sub%sub%
\subsection{Nucleation within the Quark-Meson Model}\label{su:NucleationQM}
%sub%sub%sub%

\begin{figure}
	\centering
	\includegraphics[width=.49\textwidth]{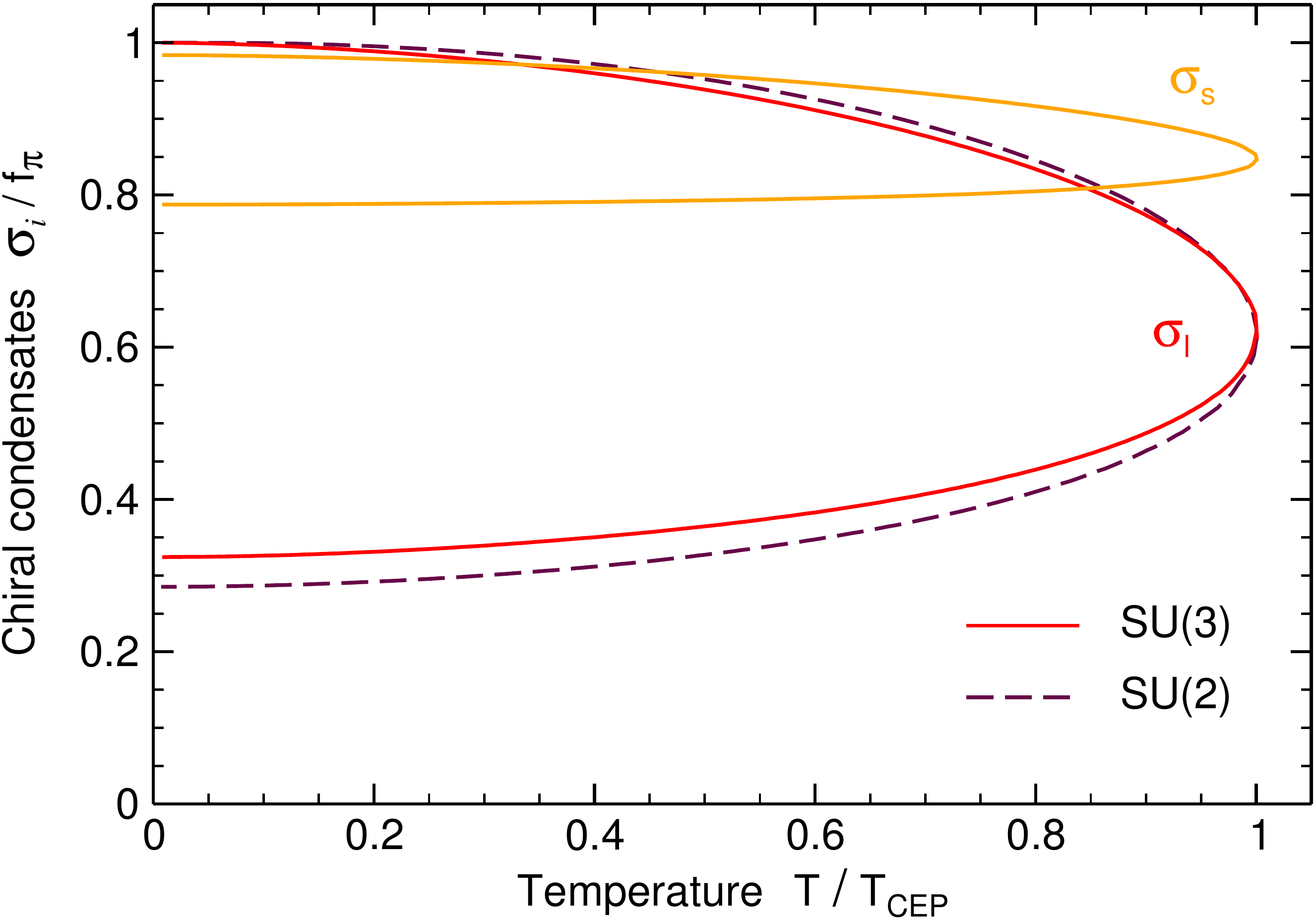}
	\hfill
	\includegraphics[width=.49\textwidth]{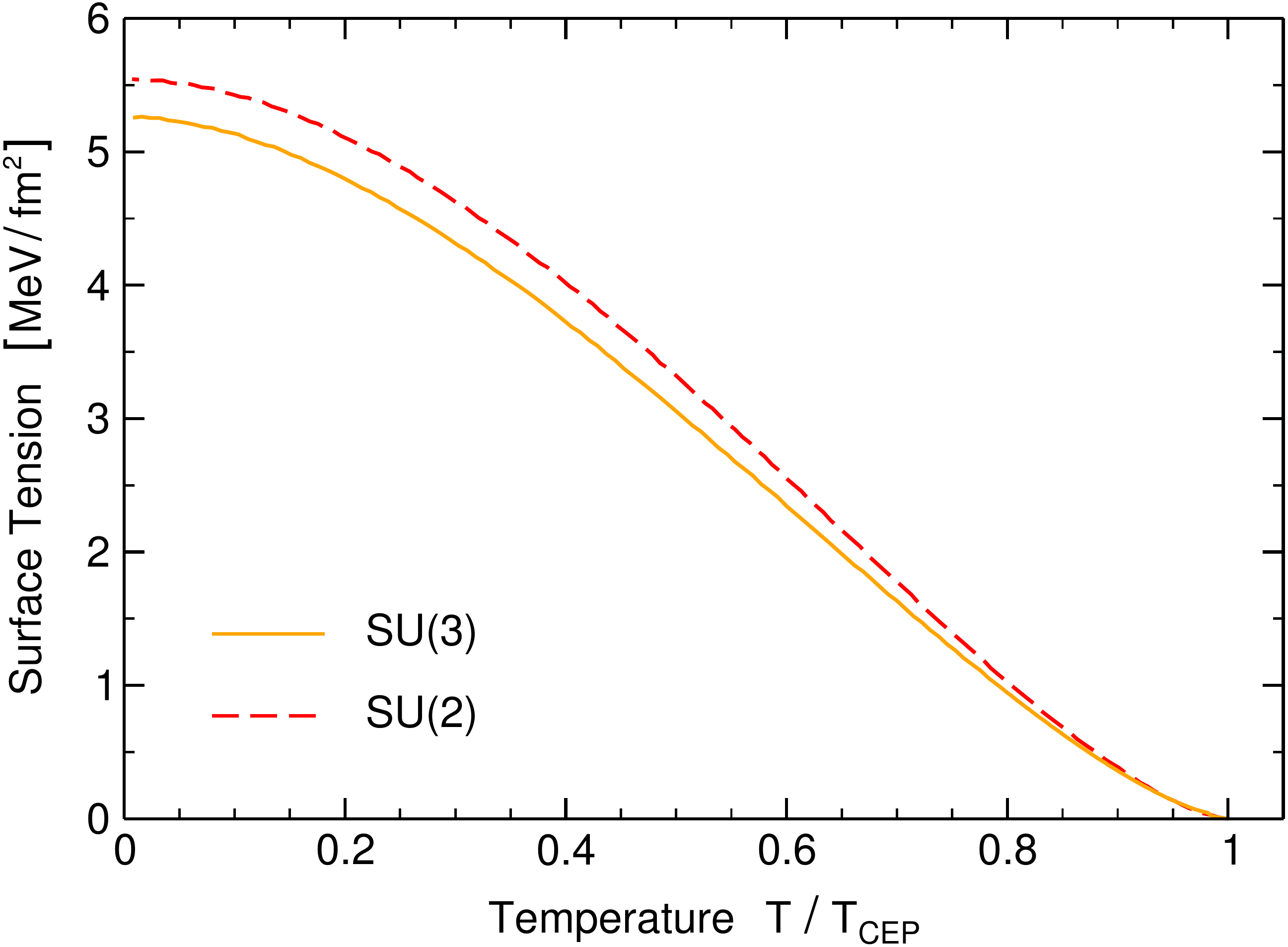}
	\caption{Left: Degenerate values of the chiral condensates $\sigma_\mrm{l}$ and $\sy$ along the first-order transition line in the two and 2+1--flavour Quark-Meson model. Right: Surface tension along the coexistence line in the two and 2+1--flavour Quark-Meson model.}
	\label{fig:QM_OPs_ST}
\end{figure}

The PQM model reduces to the Quark-Meson model that governs only chiral symmetry breaking and restoration by fixing $\blr{\Phir,\Phii}\equiv\blr{1,0}$ at all temperatures and densities. This implies that the coarse-grained free energy and the estimate of the surface tension become independent of the kinetic parameter $\kappa$ of the Polyakov-loop order parameters,
%e%e%e%
\bea
	\Sigma_\mrm{QM}^\mrm{SU(3)}\blr{T,\mu_\mrm{l},\mu_\mrm{s}} &=& \sqrt{\blr{\Delta\sigma_\mrm{l}}^2 + \blr{\Delta\sy}^2}\int_0^1\mrm{d}\xi\, \sqrt{2\tilde{\Omega}\!\blr{\xi;T,\mu_\mrm{l},\mu_\mrm{s}}}\;.
	\label{eq:sigma_QM3}
\eea
%e%e%e%
The contribution of the light quark sector is already present in the SU(2) Quark-Meson model, where the overestimate of the surface tension simply reduces to
%e%e%e%
\bea
	\Sigma_\mrm{QM}^\mrm{SU(2)}\blr{T,\mu_\mrm{l}} &=& \int_{\sigma_\mrm{l}^{\blr{1}}}^{\sigma_\mrm{l}^{\blr{2}}}\mrm{d}\sigma_\mrm{l}\, \sqrt{2{\Omega}\!\blr{\sigma_\mrm{l};T,\mu_\mrm{l}}}\;.
	\label{eq:sigma_QM2}
\eea
%e%e%e%
Hence, differences in the surface tension of the two-flavour and 2+1--flavour Quark-Meson model arise due to a possible modification of the degenerate values of the light chiral condensate, the additional dimension of the strange chiral condensate and a  possible modification of the effective potential along the straight line connecting the minima.
To investigate the impact of the values of the chiral condensates, the left part of \Fig{fig:QM_OPs_ST} shows the degenerate values that the chiral order parameters take at the first-order phase transition.
%f%f%f%
%f%f%f%
Notice that the minima merge smoothly at the critical endpoint.

One observes that adding strange quarks to the system weakens the transition in the light quark sector. This partially counteracts the increase of the prefactor in the surface tension \eq{eq:sigma_QM3} when strange quarks are taken into account.
But the upper limit on the surface tension presented on the right-hand side of \Fig{fig:QM_OPs_ST} is even smaller for the 2+1 quark flavour Quark-Meson model that in the two-flavour calculation. This is because the fermionic vacuum contribution at one-loop order
%e%e%e%
\bea
	\Omega_\qqb^\mrm{vac}\; &\sim& -\!\sum_{f=\mrm{u,d,s}} m_f^4 \ln\blr{m_f}\;,
\eea
%e%e%e%
decreases the thermodynamical potential and the height of the barrier at the first-order phase transition and each quark flavour leads to a further decrease that is even larger the heavier the quark species.

In general, \Fig{fig:QM_OPs_ST} shows the overestimate of the surface tension in the thin-wall approximation of the QM model along the first-order transition line. The temperature dependence of the 
surface tension and its value at zero temperature
is similar to the results found in \Refs{Palhares:2010be,Pinto:2012aq}, which considered the two-flavour Quark-Meson and Nambu--Jona-Lasinio models, respectively.
The upper bound of $10\,\mrm{MeV/fm}^2$ for the surface tension allows a quick hadron-quark phase conversion. The implications of this result for several physical scenarios, be it heavy-ion collisions, proto-neutron stars or the early Universe are those discussed in \Ref{Mintz:2012mz}.

%sub%sub%sub%
\subsection{Nucleation within the Polyakov--Quark-Meson Model}\label{su:NucleationPQM}
%sub%sub%sub%

With the Polyakov-loop extension of the Quark-Meson model, the system contains another order parameter, the Polyakov loop and the effective potential gets an additional contribution, the Polyakov-loop potential $\mc{U}$. 
%e%e%e%
\bea
	\Sigma\blr{\mc{R}} &=& h\int_0^1\mrm{d}\xi\, \sqrt{2\tilde{\Omega}\!\blr{\xi;\mc{R}}}\;,
	\label{eq:sigma_thinwall_straight_line}
\eea
%e%e%e%
with
%e%e%e%
\bea
	h^2&=& \blr{\Delta\sigma_\mrm{l}}^2 + \blr{\Delta\sy}^2 + \blr{\kappa\Delta\Phir}^2+\blr{\kappa\Delta\Phii}^2\;,
	\label{eq:definition_h2}
\eea
%e%e%e%
and the straight-line approximation
%e%e%e%
\bea
	\phi_i &=& \xi \phi_i^{\blr{1}} + \blr{1-\xi}\phi_i^{\blr{2}}
\eea
%e%e%e%
with $0\leq\xi\leq 1$, see further \Ref{Mintz:2012aua}.
Therefore, the Polyakov loop adds an additional  contribution to the distance between the degenerate minima at the phase transition, i.e., to the factor $h$ in \Eq{eq:definition_h2}.
Another effect of the Polyakov-loop extension on this quantity can be the modification of the values of the chiral condensates at the minima compared to the Quark-Meson model.

\begin{figure}%[t]
	\centering
	\includegraphics[width=.49\textwidth]{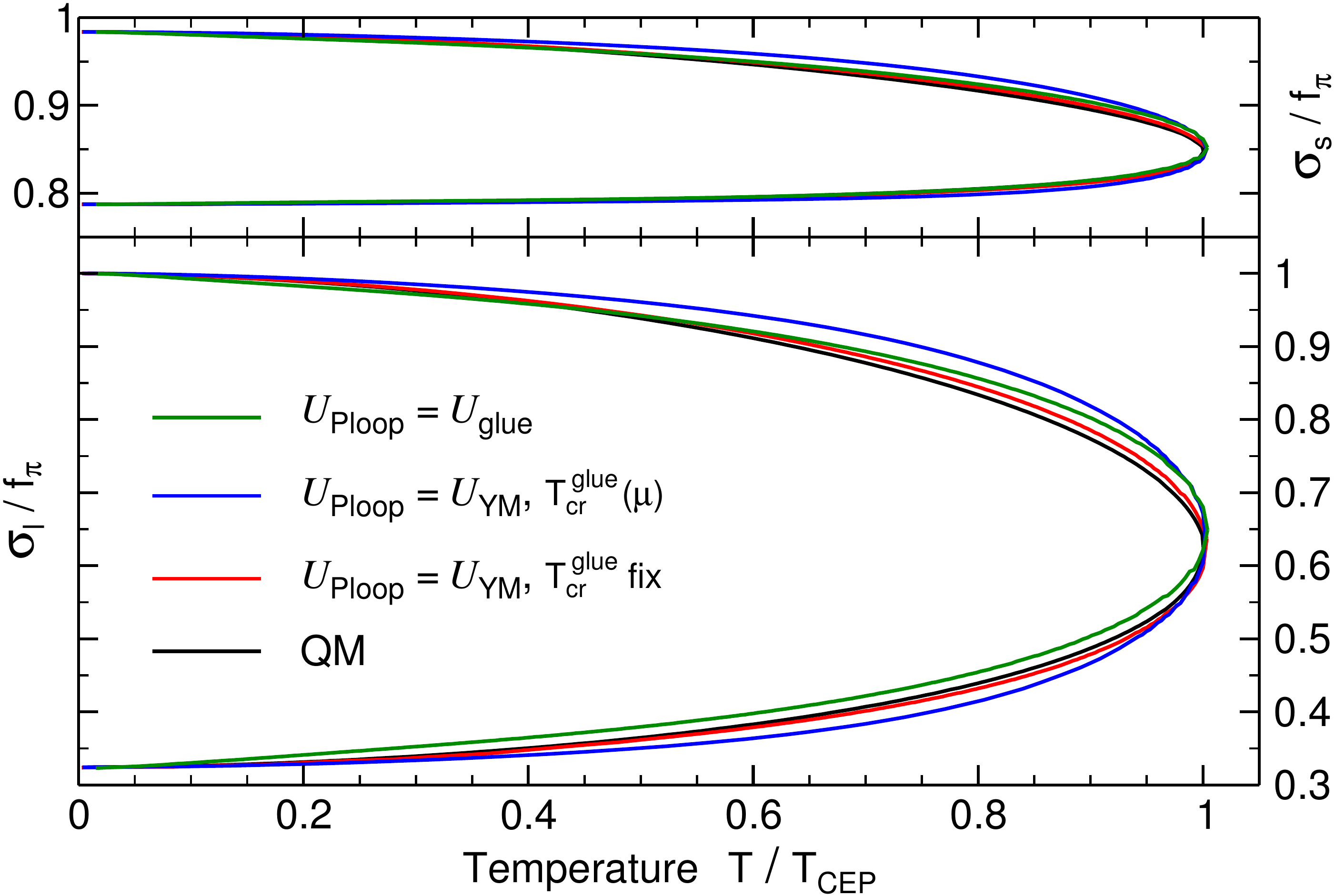}
	\hfill
	\includegraphics[width=.49\textwidth]{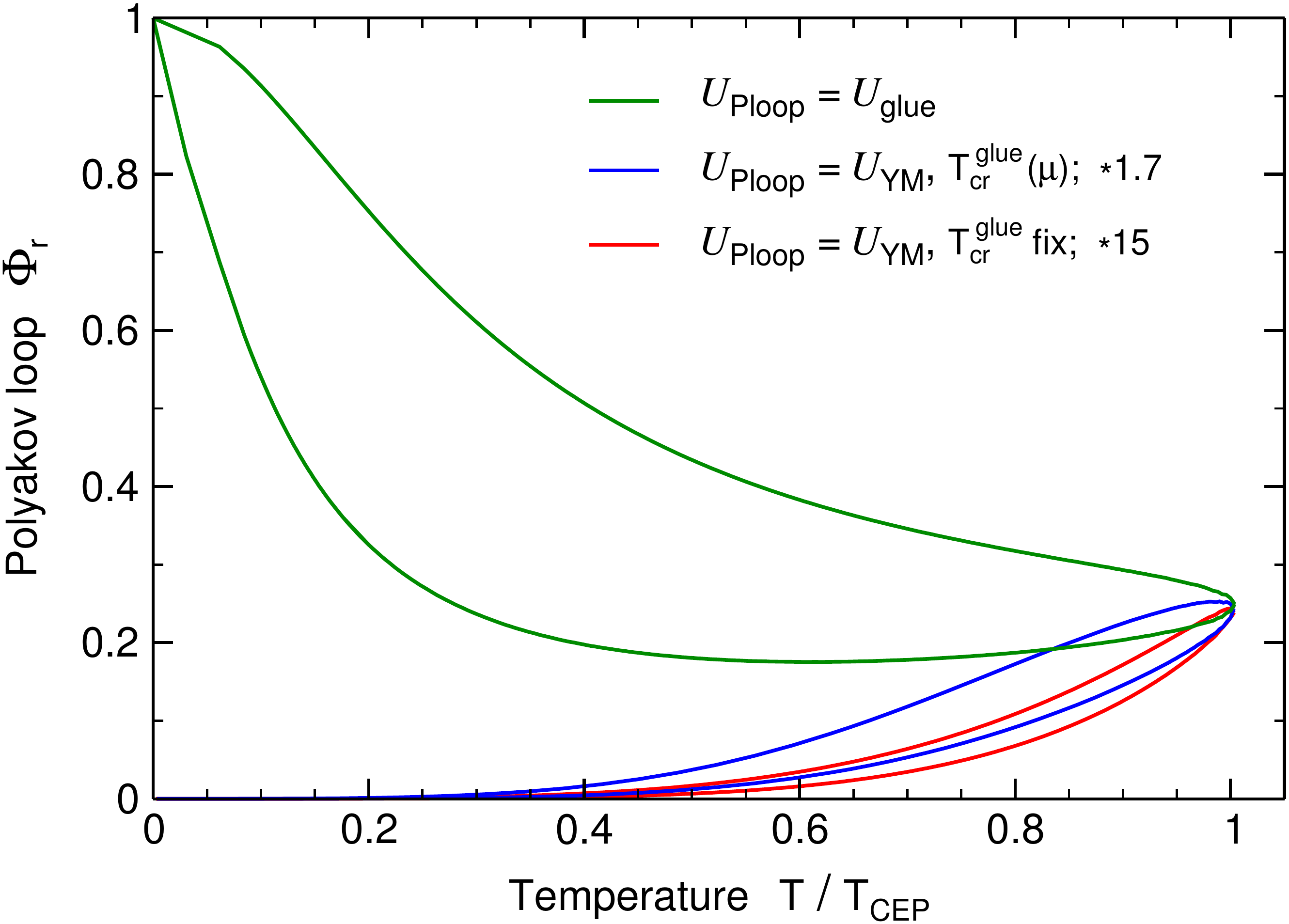}
	\caption{Degenerate values of the chiral condensates $\sigma_\mrm{l}$ and $\sy$ (left) and the Polyakov loop $\Phir$ (right) along the first-order transition line. Compared are the results of the PQM model with the Yang-Mills Polyakov-loop potential with constant and density dependent critical scale and with the unquenched Polyakov-loop potential with constant $T_\mrm{cr}^\mrm{glue}$. Note that the results for the Polyakov loop are scaled as given in the label so that they start out at similar values at the critical endpoint.	}
	\label{fig:PQM_OPs_coexistence}
\end{figure}

The left part of \Fig{fig:PQM_OPs_coexistence} shows how the values of the chiral condensates of the degenerate minima at the phase transition are altered by the coupling to the Polyakov loop. This result shows a slight dependence on whether and what kind of backcoupling of the quarks on the quenched Polyakov-loop potential is considered. With the Yang-Mills Polyakov-loop potential with a constant transition scale the values of the chiral condensates are hardly altered compared to the Quark-Meson model result. Lowering the critical temperature of this potential with increasing density increases the gap between the values of the chiral condensates in the degenerate minima. This is not the case when the unquenched Polyakov-loop potential with constant critical scale is considered but it shifts the values of the condensates somewhat to larger values. The latter observation can be attributed to the close link between the (de)confinement and chiral transitions.
%f%f%f%
%f%f%f%

The values that the Polyakov loop takes in the degenerate minima at the transition are much more sensitive to the kind of Polyakov-loop potential as is shown on the right of \Fig{fig:PQM_OPs_coexistence}. As discussed in the discussion of \Fig{fig:orderparsT10_YMglue}, the (de)confinement and chiral transitions of the light quarks remain linked in the low temperature and high density region of the phase diagram with the unquenched Polyakov-loop potential. The opposite behaviour occurs with the pure Yang-Mills Polyakov-loop potential with which the Polyakov loop takes only very small values at temperatures far below the critical scale of the Polyakov-loop potential. An intermediate behaviour for the Polyakov loop is observed with the chemical potential dependence of the transition temperature of the Polyakov-loop potential as in \Ref{Schaefer:2007pw}. Close to the critical endpoint the Polyakov loop takes at the phase transition values of the same order as with the unquenched Polyakov-loop potential but nevertheless the (de)confinement transition decouples from the chiral phase transition of the light quarks at even smaller temperatures.

\begin{figure}%[h]
	\includegraphics[width=0.6\textwidth]{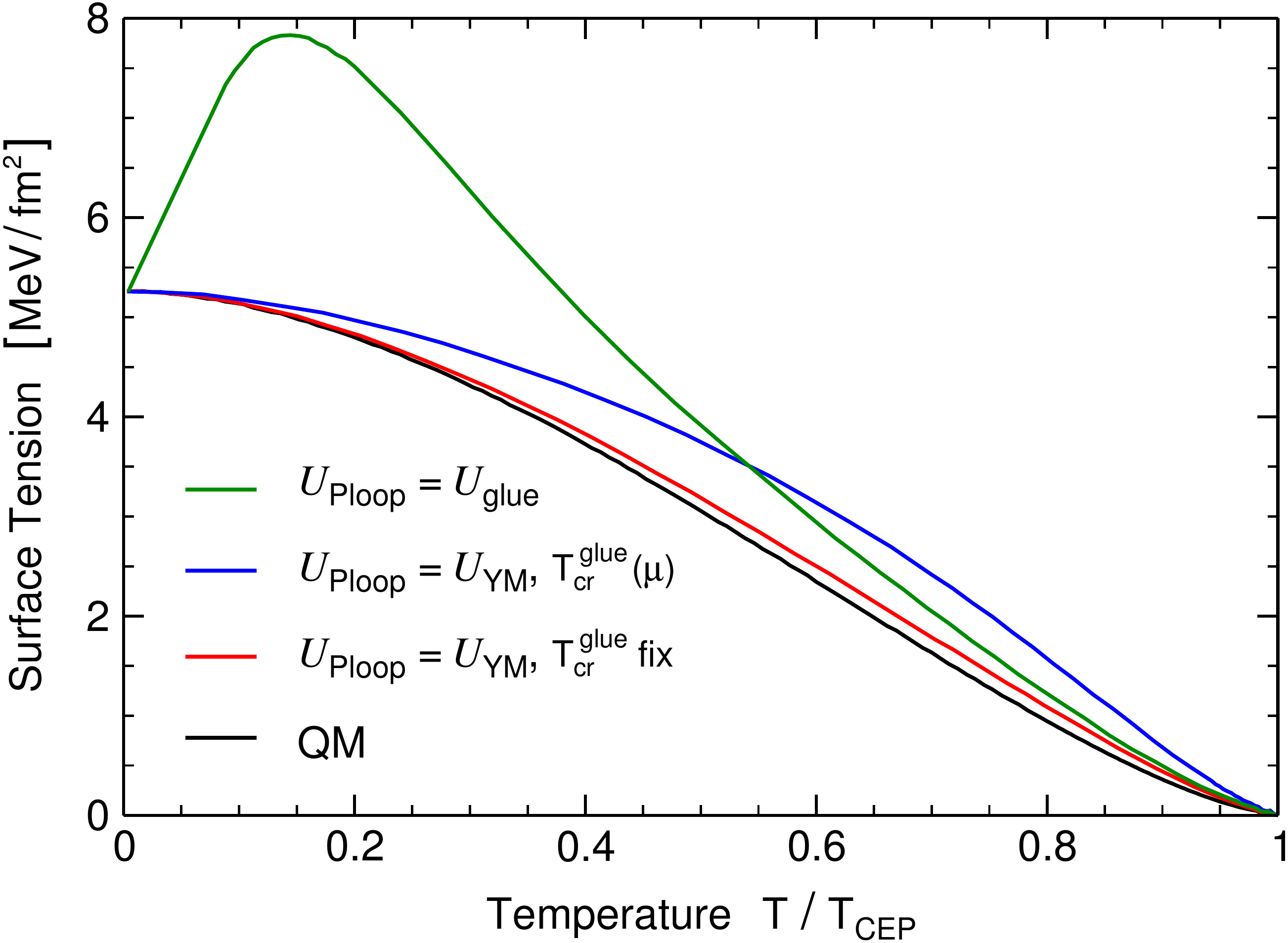}
	\hskip-10ex
	\begin{minipage}[b]{0.5\textwidth}
		\caption{\label{fig:SurfaceTension_glueYMfixmu}Surface tension along the coexistence line. Compared are the results of the PQM model with the Yang-Mills Polyakov-loop potential with constant and density-dependent critical scale and with the unquenched Polyakov-loop potential with constant $T_\mrm{cr}^\mrm{glue}$.}
	\end{minipage}
\end{figure}

These results are obtained with the polynomial-logarithmic parametrisation of the Polyakov-loop potential with the best-fit parameters of \Sec{sec:Resultsmu0}.
Figure \ref{fig:SurfaceTension_glueYMfixmu} presents the corresponding results of the surface tension. The kinetic parameter $\kappa$ of the Polyakov loop is adjusted via the pure gauge surface tension to $\kappa\simeq0.818\,T_0$.
%f%f%f%
%f%f%f%
The Polyakov-loop extension of the pure chiral Quark-Meson model leads to an increase of the surface tension. Nevertheless, the zero-temperature limit is the same since gluon excitations are independent of the quark chemical potential in this limit, $\mc{U}\blr{\Phir,\Phii;T=0}=0$ $\forall\blr{\Phir,\Phii}$.
Except for very close to the critical endpoint, $\Delta\sigma_\mrm{l}$ is much lager than $\Delta\Phir$, $\Delta\Phii$ and $\Delta\sy$ so that the order of magnitude of the surface tension is set by the two-flavour Quark-Meson model. The maximum in the surface tension with the unquenched Polyakov-loop potential is the result of the increasing strength of the (de)confinement transition that comes along with the stronger transition in the chiral sector at larger chemical potentials and the tendency that a (de)confinement crossover transition sets in at chemical potentials slightly below the chiral first-order transition.

%s%s%s%s%s%s%s%
\section{Conclusions}\label{sec:Conclusions}
%s%s%s%s%s%s%s%

In this work, thermodynamical properties and the phase structure of strongly-inter\-acting matter have been analysed in an improved framework of the Polyakov-loop--extended Quark-Meson model with 2+1 constituent quark flavours.
We applied the results of \Ref{Haas:2013qwp} that an appropriate rescaling of the temperature can mimic the effect of the quark backreaction on the gauge sector. This offers a simple and systematic approach to improve the Polyakov-loop potential from a pure gauge potential to the unquenched glue potential in full QCD.
Moreover, as another step beyond usual mean-field analyses not only quark quantum fluctuations at one-loop have been considered but the contribution of thermal meson fluctuations to thermodynamics was also taken into account.

With the parameters adjusted in order to reproduce lattice results at zero quark densities this setting constitutes an adequate framework to investigate the phase structure of strongly-interacting matter at non-zero quark density.
The present work presents the first application of the unquenched Polyakov-loop potential at non-zero net quark density and explored its impact. 
Including the quark backreaction on the gauge sector showed to have a big impact on the interrelation of the chiral and (de)confinement transitions at non-vanishing chemical potential and these transitions remained linked even in the high quark-density and small temperature region of the phase diagram. This result is obtained ignoring any density dependence of the quark backreaction. So, in order to confirm or reject this trend it remains for future work to investigate a medium dependence of the unquenching of the gauge sector.

The comparison of the curvature of the phase transition line at non-zero baryon chemical potential with recent lattice data and observations of the latest hadron chemical equilibrium in relativistic heavy-ion collisions showed an appreciable agreement.
Testing the effective model with its parameters adjusted to provide a good description of lattice data at zero density in this fashion is crucial to understand whether it can provide a reliable description of the phase structure of the strong interaction in general.

The careful circumvention of the fermion sign problem in a way that preserved the solutions of the equation of motion as minima of the effective potential in \Ref{Mintz:2012mz} allowed to derive a formalism to study the homogeneous nucleation of bubbles in a first-order phase transition. An upper limit of the surface tension for bubble nucleation in the thin-wall approximation has been calculated. 
Here, we found that the Polyakov--Quark-Meson model with 2+1 quark flavours yields results similar to those of the two-flavour Quark-Meson model, so that the influence of both, the strange quark and the Polyakov loop at low temperatures is small. The conservative upper bound of $10\,\mrm{MeV/fm}^2$ found for the surface tension allows a quick hadron-quark phase conversion.

Concerning possible applications, an ingredient of investigations of quarkonium suppression in a quark-gluon plasma formed in heavy-ion collisions is the expansion of the plasma itself \cite{Gossiaux:2009mk,Marty:2012vs,Marty:2013ita,Marty:2014zka}. Here, the temperature and density dependence of quark and meson masses found in the presented framework can serve as an input.

\ack
We thank Eduardo S.~Fraga, Hubert Hansen, Bruno W.~Mintz, Mario Mitter, Jan M.~Pawlowski, Tina K.~Herbst and Andreas Zacchi for valuable discussions and collaboration on related topics. RS acknowledges support by  the Helmholtz International Center for FAIR (HIC for FAIR) and the Service pour la science et la technologie pr\`es l'Ambassade de France en Allemagne and Campus France.

\appendix

\section{Polyakov-loop potentials that respect the SU(3) group volume}\label{app:PloopPots}

The potential of the Polyakov loop, $\mc{U}\!\blr{\Phir,\Phii;T}$ should mimic a background of gluons and controls the dynamics of the Polyakov loop.
First computations of the effective potential in gauge theories have been performed in the 80's at asymptotically high temperatures using perturbation theory \cite{Weiss:1980rj,Gross:1980br,Weiss:1981ev} and in the strong-coupling limit on a lattice \cite{Svetitsky:1985ye}.
In recent years, the non-perturbative Polyakov-loop potential has been studied using various different approaches \cite{Braun:2007bx,Marhauser:2008fz,Braun:2009gm,Braun:2010cy,Pawlowski:2010ht,Sasaki:2012bi,Ruggieri:2012ny,Dumitru:2012fw,Diakonov:2012dx,Fukushima:2012qa,Reinhardt:2012qe,Kashiwa:2012td,Fister:2013bh,Fischer:2013eca,Herbst:2015ona}. First-principle calculations of the potential are performed using different functional methods, mainly the FRG approach but also Dyson-Schwinger equations and the two particle irreducible (2PI) approach \cite{Braun:2007bx,Marhauser:2008fz,Braun:2010cy,Fister:2013bh,Fischer:2013eca,Herbst:2015ona}.
In \Refs{Braun:2009gm,Pawlowski:2010ht} the Polyakov-loop potential in two flavour QCD in the chiral limit has been analysed. This computation includes the full backcoupling of the matter sector on the propagators of the gauge degrees of freedom via dynamical quark-gluon interactions \cite{Gies:2002hq,Braun:2006jd,Braun:2008pi,Braun:2014ata}.\\
A much simpler way to obtain an effective Polyakov-loop potential  $\mc{U}\!\blr{\Phi,\Phib;T}$ is to construct a potential that respects all given symmetries and contains the spontaneous breaking of Z(3) symmetry if the system is in the deconfined phase \cite{Svetitsky:1982gs,Svetitsky:1985ye,Banks:1983me}.
The simplest terms that lead to a real potential, respect centre symmetry and are able to describe spontaneous symmetry breaking are a combination of terms of second and forth order $\mc{U} \sim p_2\Phi\Phib + p_4\blr{\Phi\Phib}^2$. The combination $\Phi\Phib$ is also invariant under $U\!\blr{1}$ transformations. But the potential governing centre symmetry should not introduce any additional symmetries. Therefore, the potential also has to contain terms that break the global symmetry down from $O\!\blr{2}$ to $Z\!\blr{3}$. The simplest real term that is invariant under $Z\!\blr{3}$ but breaks $U\!\blr{1}$ and is symmetric under charge conjugation is $\blr{\Phi^3+\Phib^3}$.
The polynomial of the above-mentioned terms form the minimal content of a Polyakov-loop potential \cite{Pisarski:2000eq,Ratti:2005jh}, $\mc{U} \sim p_2\!\blr{t} \Phi\Phib + p_3 \blr{\Phi^3+\Phib^3} + p_4 \blr{\Phi\Phib}^2$.
The coefficient $p_2$ has to be temperature dependent to realise the transition to a phase where $Z\!\blr{3}$ symmetry is spontaneously broken.
For later use, this temperature dependence is written in terms of a reduced temperature $t=\blr{T-T_\mrm{c}}/T_\mrm{c}$ where $T_\mrm{c}=T_0$ is in this case the transition temperature of the Polyakov-loop potential.
A negative cubic term forces the nature of the transition to be of first order while it would be of second order otherwise.
The coefficient of the forth-order term has be positive so that the potential is bounded from below for large $\Phi$ and $\Phib$.
The ansatz for the Polyakov-loop potential can be enhanced by including the term that arises if one integrates out the SU(3) group volume in the generating functional for the Euclidean action. This integration can be performed via the so-called Haar measure and takes the form of a Jacobian determinant. Its logarithm adds as an effective potential to the action in the generating functional.
This function already breaks the $U\blr{1}$ symmetry and with a positive coefficient the logarithm bounds the potential from below for large $\Phi$ and $\Phib$, so that one can drop the cubic and forth-order terms of the polynomial while the kinetic part $\sim \Phi\Phib$ remains \cite{Fukushima:2003fw,Roessner:2006xn},
%e%e%e%
\bea
	\hskip-5ex \frac{\mc{U}_\mrm{log}\blr{\Phi,\Phib,t}}{T^4} &=& p_2\!\blr{t} \Phi\Phib + l\!\blr{t} \ln \left[ 1 - 6 \Phi\Phib + 4 \blr{\Phi^{3}+\Phib^{3}} - 3 \blr{ \Phi\Phib }^{2}\right]\;.
	\label{eq:PloopPot_log}
\eea
%e%e%e%
This potential is also qualitatively consistent with the leading order result of the strong-coupling expansion \cite{Gross:1983ju,Langelage:2010yr}.
An additional feature of the logarithmic term is that the potential diverges for $\Phi,\Phib\to1$ thus limiting the Polyakov loop to be always smaller than one, reaching this value only asymptotically as $T\to\infty$. This is consistent with the relation of the Polyakov loop to the free energy of a static quark-antiquark pair.\\
Reference \cite{Lo:2013hla} went beyond a minimal content for the Polyakov-loop potential and kept the higher-order terms of the polynomial parametrisation of the Polyakov-loop potential and added the logarithmic term to consider the group volume additionally,
%e%e%e%
\bea
	\frac{\mc{U}_\mrm{polylog}\blr{\Phi,\Phib,t}}{T^4} &=& p_2\!\blr{t} \Phi\Phib + p_3\!\blr{t} \blr{\Phi^3+\Phib^3} + p_4\!\blr{t} \blr{\Phi\Phib}^2 + \nonumber\\
	&& + \,l\!\blr{t} \ln \left[ 1 - 6 \Phi\Phib + 4 \blr{\Phi^{3}+\Phib^{3}} - 3 \blr{ \Phi\Phib }^{2}\right]\;.
	\label{eq:PloopPot_polylog}
\eea
%e%e%e%

\begin{figure}%[h]
	\includegraphics[width=0.6\textwidth]{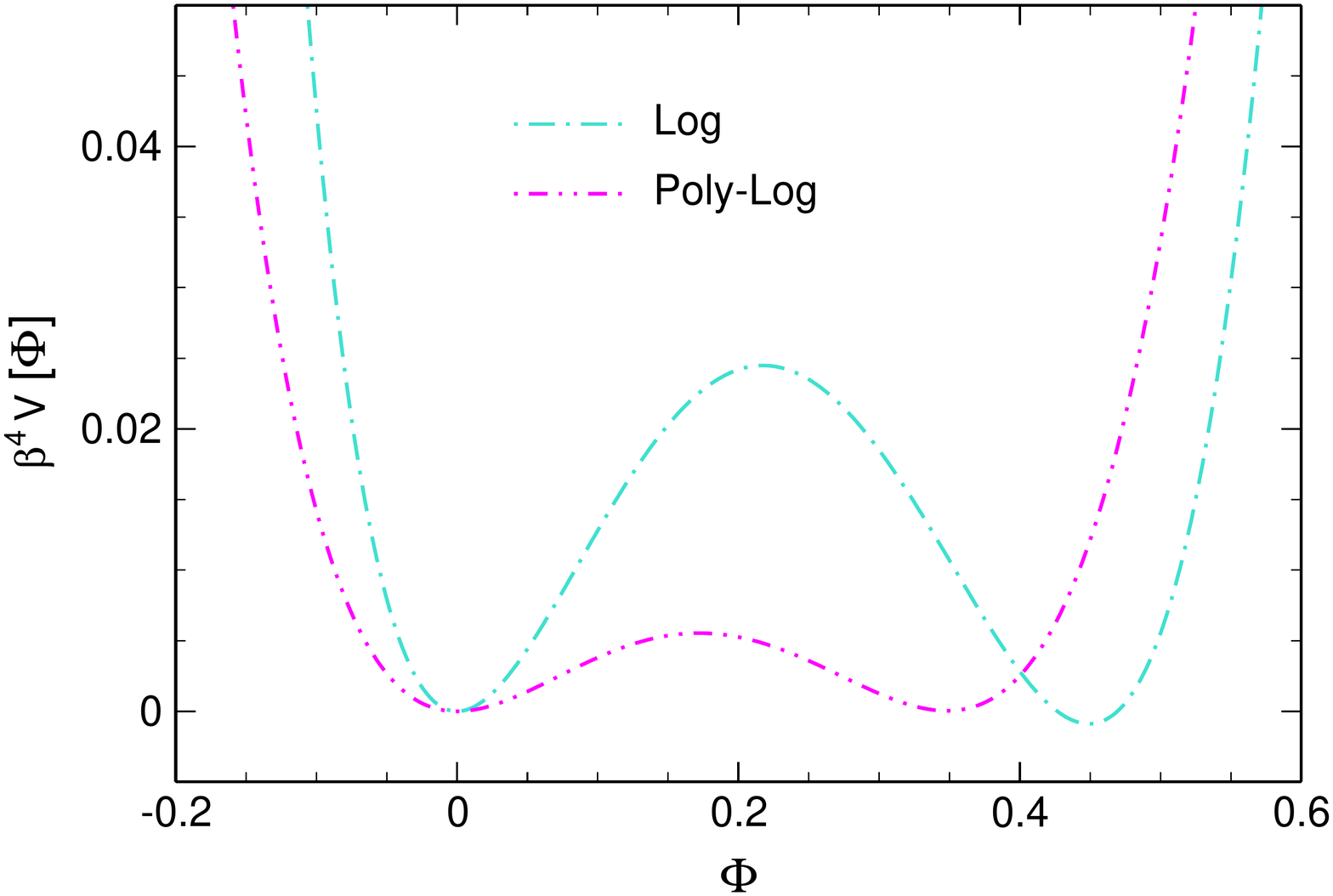}
	\hskip-10ex
	\begin{minipage}[b]{0.5\textwidth}
		\caption{\label{fig:PolPots_t0}Polyakov-loop potentials at their transition scale $t=0$.}
	\end{minipage}
\end{figure}

In $\mc{U}_\mrm{log}$ the temperature dependence of the coefficients is parametrised as a polynomial
%e%e%e%
\bea
	c\blr{t} &=& \sum_{n} \frac{C_n}{\blr{1+t}^n}\;,
\eea
%e%e%e%
where $t=\blr{T-T_\mrm{c}}/T_\mrm{c}$ defines a reduced temperature and $T_\mrm{c}=T_0$ is in this case the transition temperature of the Polyakov-loop potential.
The coefficients of the polynomial-logarithmic parametrisation were defined in \Ref{Lo:2013hla} in a more complex way,
%e%e%e%
\bea
	p_i\blr{t} &=& \left[P_0^{\blr{i}} + \frac{P_1^{\blr{i}}}{1+t} +  \frac{P_2^{\blr{i}}}{\blr{1+t}^2}\right]\Bigg/\left[1 + \frac{P_3^{\blr{i}}}{1+t} +  \frac{P_4^{\blr{i}}}{\blr{1+t}^2}\right]
\eea
%e%e%e%
and
%e%e%e%
\bea
	l\!\blr{t} &=& \frac{L_0}{\blr{1+t}^{L_1}} \left[ 1- e^{{L_2}/{\blr{1+t}^{L_3}}} \right]\;.
\eea
%e%e%e%
The number of independent parameters of the parametrisations can be reduced by imposing some general constraints. One condition is that the Polyakov-loop variables approach unity for large temperatures. A necessary condition for the expectation values of the Polyakov loops is $\pd\mc{U}/\pd\Phi = \pd\mc{U}/\pd\Phib = 0$. To fix further parameters one applies and restricts the Polyakov-loop potential to pure gauge (Yang-Mills) theory.
The absence of dynamical quarks restricts the Polyakov-loop variable to be real, $\Phi=\Phib$. At the transition scale of the Polyakov-loop potential $T_0$, a first-order phase transition is required. If a reliable prediction of the value of the Polyakov loop at the phase transition would be available it would further constrain the parameters. Furthermore, a gas of $\Nc-1$ non-interacting, massless gluons should approach its Stefan-Boltzmann limit as $T\to\infty$.
The remaining open parameters are determined in \Refs{Scavenius:2002ru,Ratti:2005jh,Roessner:2006xn,Lo:2013hla} by fitting both the lattice data for pressure, entropy density and energy density and the evolution of the Polyakov loop $\langle \Phi \rangle$ on the lattice in pure gauge theory. Reference~\cite{Lo:2013hla} adjusted their parameters in addition to lattice data of the longitudinal and transverse Polyakov-loop susceptibilities.
The different parameter sets are summarised in \Tab{tab:Ploop_pot_params}.
%t%t%t%
\begin{table}
	\caption{Parameters of the different parametrisations of the Polyakov-loop potential for fits to the lattice Yang-Mills simulations \cite{Kaczmarek:2002mc,Boyd:1996bx} and \cite{Lo:2013etb,Borsanyi:2012ve}.}
	\begin{center}
		\begin{tabular}{lcccccc}
			\hline\hline\vspace{-0.3cm}\\
			        							& $P_0$	& $P_1$	& $P_2$	& $L_3$	\\
			Log \cite{Roessner:2006xn} 		& -1.755 	& 1.235 	& -7.6 	&-1.75	\\
			\vspace{-0.3cm}\\\hline\vspace{-0.3cm}\\
			Poly-Log \cite{Lo:2013hla}		& $P_0^{\blr{2}}$	& $P_1^{\blr{2}}$	& $P_2^{\blr{2}}$	& $P_3^{\blr{2}}$	& $P_4^{\blr{2}}$	\\
										& 22.07			& -75.7			& 45.03385		& 2.77173		& 3.56403	\\
										& $P_0^{\blr{3}}$	& $P_1^{\blr{3}}$	& $P_2^{\blr{3}}$	& $P_3^{\blr{3}}$	& $P_4^{\blr{3}}$	\\
										& -25.39805		& 57.019			& -44.7298		& 3.08718		& 6.72812	\\
										& $P_0^{\blr{4}}$	& $P_1^{\blr{4}}$	& $P_2^{\blr{4}}$	& $P_3^{\blr{4}}$	& $P_4^{\blr{4}}$	\\
										& 27.0885			& -56.0859			& 71.2225		& 2.9715		& 6.61433	\\
										& $L_0$			& $L_1$			& $L_2$			& $L_3$	\\
										& -0.32665		& 5.85559			& -82.9823				& 3.0\\\vspace{-0.3cm}\\					
			\hline\hline
		\end{tabular}
	\end{center}
	\label{tab:Ploop_pot_params}
\end{table}
%t%t%t%
Results for quantities that characterise the Polyakov-loop potential are given in \Tab{tab:Ploop_pot_data}.
%t%t%t%
\begin{table}
	\caption{Numbers characterising the Polyakov-loop potential. It should be $\Phi_{T\to\infty}=1$, $\blr{\mc{U}/p_\mrm{SB}}_{T\to\infty}=1$ and $\mc{U}\blr{\Phi_{t=0}}=0$.}
	\begin{center}
		\begin{tabular}{lcrcc}
			\hline\hline\vspace{-0.3cm}\\
			        		& $\Phi_{t=0}$	& $\mc{U}/T^4\blr{\Phi_{t=0}}$	& $\Phi_{T\to\infty}$	& $\blr{\mc{U}/p_\mrm{SB}}_{T\to\infty}$	\\
			\vspace{-0.3cm}\\\hline\vspace{-0.3cm}\\
			Log		& 0.449		& $-8.84\times10^{-4}$		& 1.0				& 1.0		\\
			Poly-Log	& 0.348 		& $-1.18\times10^{-4}$		& 1.0				& 0.933	\\
			\vspace{-0.3cm}\\	
			\hline\hline
		\end{tabular}
	\end{center}
	\label{tab:Ploop_pot_data}
\end{table}
%t%t%t%
An apparent difference is that the polynomial-logarithmic parametrisation reaches only $\sim93\,\%$ of the high-temperature expectation.
Figure \ref{fig:PolPots_t0} compares the form of the potentials at their critical scale, i.e.~at $t=0$.
It shows another important difference between the different parametrisations. The barrier between the two minima at the transition temperature differs in its width and height between the different parametrisations. This affects nucleation in Yang-Mills theory and we discussed in \Ref{Mintz:2012mz} how one can partially account for the impact of these different barriers on nucleation within the Polyakov--Quark-Meson model and how they influence the result of the surface tension.

Rewritten as functions of the real variables $\Phir=\blr{\Phi+\Phib}/{2}$ and $\Phii=\blr{\Phi-\Phib}/{2}$, the parametrisations of the Polyakov-loop potential are
%e%e%e%
\bea
	\hskip-15ex \frac{\mc{U}_\mrm{log}\blr{\Phir,\Phii,t}}{T^4} &=& p_2\!\blr{t} \blr{\Phir^2+\Phii^2} +\non\\
	&&+l\!\blr{t} \ln \left[ 1 - 6\blr{\Phir^2 + \Phii^2} + 8 \blr{ \Phir^3 - 3\Phir\Phii^2} - 3 \blr{\Phir^2 + \Phii^2}^2\right]\;,\label{eq:LogPloopPot_ri} \\
	\hskip-15ex \frac{\mc{U}_\mrm{polylog}\blr{\Phir,\Phii,t}}{T^4} &=& p_2\!\blr{t} \blr{\Phir^2+\Phii^2} + 2 p_3\!\blr{t} \blr{\Phir^2  -3\Phir\Phii^2} + p_4\!\blr{t} \blr{\Phir^2+\Phii^2}^2 +\non \\
	&&+ l\!\blr{t} \ln \left[ 1 - 6\blr{\Phir^2 + \Phii^2} + 8 \blr{ \Phir^3 - 3\Phir\Phii^2} - 3 \blr{\Phir^2 + \Phii^2}^2\right]\;.\label{eq:PolyLogPloopPot_ri}
\eea
%e%e%e%

\vskip5ex

\bibliographystyle{utphys}
\bibliography{../references}

\end{document}